\documentclass[12pt]{article}

\usepackage{amssymb}
\usepackage{amsmath}
\usepackage{amsthm}
\usepackage{color}
\usepackage{comment}
\usepackage{enumitem}		
\usepackage{geometry}		
\usepackage{graphicx}
\usepackage{authblk}
\usepackage{amssymb}
\usepackage{listings}
\usepackage{xcolor}
\usepackage{caption}
\usepackage{color}
\usepackage{algorithm}
\usepackage{algpseudocode}

\usepackage{subcaption}

\makeatletter
	\@addtoreset{equation}{section}
\makeatother

\let\originalleft\left
\let\originalright\right
\renewcommand{\left}{\mathopen{}\mathclose\bgroup\originalleft}
\renewcommand{\right}{\aftergroup\egroup\originalright}

\newlist{romanlist}{enumerate}{3}
\setlist[romanlist]{label=\roman*),ref=(\roman*)}

\begin{document}

\newcommand{\cF}{\mathcal{F}}
\newcommand{\cP}{\mathcal{P}}
\newcommand{\cR}{\mathcal{R}}
\newcommand{\cS}{\mathcal{S}}
\newcommand{\cT}{\mathcal{T}}
\newcommand{\cW}{\mathcal{W}}
\newcommand{\ee}{\varepsilon}
\newcommand{\rD}{{\rm D}}
\newcommand{\re}{{\rm e}}

\newtheorem{theorem}{Theorem}[section]
\newtheorem{corollary}[theorem]{Corollary}
\newtheorem{lemma}[theorem]{Lemma}
\newtheorem{proposition}[theorem]{Proposition}

\theoremstyle{definition}
\newtheorem{definition}{Definition}[section]


\title{
The bifurcation structure within robust chaos for two-dimensional piecewise-linear maps.
}

\author[1]{I.~Ghosh}
\author[1]{R.I.~McLachlan}
\author[1]{D.J.W.~Simpson}

\affil[1]{School of Mathematical and Computational Sciences\\
Massey University\\
Colombo Road, Palmerston North, 4410\\
New Zealand}

\maketitle


\begin{abstract}
We study two-dimensional, two-piece, piecewise-linear maps having two saddle fixed points.
Such maps reduce to a four-parameter family and are
well known to have a chaotic attractor throughout open regions of parameter space.
The purpose of this paper is to determine
where and how this attractor undergoes bifurcations.
We explore the bifurcation structure numerically by
using Eckstein's greatest common divisor algorithm
to estimate from sample orbits the number of connected components in the attractor.
Where the map is orientation-preserving the numerical results agree with formal results
obtained previously through renormalisation.
Where the map is orientation-reversing or non-invertible the same renormalisation
scheme appears to generate the bifurcation boundaries,
but here we need to account for the possibility of some stable low-period solutions.
Also the attractor can be destroyed in novel heteroclinic bifurcations (boundary crises)
that do not correspond to simple algebraic constraints on the parameters.
Overall the results reveal a broadly similar component-doubling bifurcation structure
in the orientation-reversing and non-invertible settings,
but with some additional complexities.
\end{abstract}

\section{Introduction}
\label{sect:intro}

Complex dynamics are caused by nonlinearity,
and many physical systems involve an extreme form of nonlinearity: a switch.
Examples include engineering systems with relay or on/off control
which involve distinct modes of operation \cite{Bu01,Li03,Ts84}.
Mathematical models of systems with switches are naturally piecewise-smooth,
where different pieces of the equations correspond to different positions of a switch.

Piecewise-smooth maps arise as discrete-time models
of systems with switches \cite{IbCa11,PuSu06}
and as Poincar\'e maps or stroboscopic maps of continuous-time models \cite{DiBu08}.
For multi-dimensional piecewise-smooth maps
there are many {\em abstract} ergodic theory results
stating that if an attractor satisfies certain properties (e.g.~expansion)
then it is chaotic in a certain sense
(e.g.~has an SRB measure) \cite{Bu99,Pe92,Ry04,Ts01,Yo85}.
This paper in contrast provides {\em explicit} results for a physically-motivated family of maps.
We study the four-parameter family
\begin{align}
\label{eq:BCNF2}
f_{\xi}(x, y) = \begin{cases}
\begin{bmatrix}\tau_Lx + y + 1 \\ -\delta_Lx \end{bmatrix}, & x \leq 0, \\ \begin{bmatrix}\tau_Rx + y +1 \\ -\delta_Rx \end{bmatrix}, & x \ge 0,
\end{cases}
\end{align}
where $\xi = (\tau_L,\delta_L,\tau_R,\delta_R) \in \mathbb{R}^4$.
This is known as the two-dimensional border-collision normal form (BCNF) \cite{NuYo92}.
This family provides leading-order approximations
to border-collision bifurcations where a fixed point of a piecewise-smooth map
collides with a switching manifold as parameters are varied \cite{Si16}.
As such \eqref{eq:BCNF2} can be used to capture the change in dynamics
as parameters are varied to pass through border-collision bifurcations.
In this regards it has been applied to
mechanical oscillators with stick-slip friction \cite{SzOs09},
DC/DC power converters \cite{ZhMo06b},
and various models in economics \cite{PuSu06}.
We stress that unlike Poincar\'e maps of smooth invertible flows,
the piecewise-smooth maps arising from these applications are mostly non-invertible. 
We distinguish four generic cases based on the signs of
the determinants $\delta_L$ and $\delta_R$:
\begin{itemize}
    \item With $\delta_L>0$ and $\delta_R>0$ \eqref{eq:BCNF2} is orientation-preserving. This is the classical scenario considered in the `robust chaos' paper of
    Banerjee {\em et al} \cite{BaYo98}.
    \item With $\delta_L<0$ and $\delta_R<0$ \eqref{eq:BCNF2} is orientation-reversing. Misiurewicz \cite{Mi80} used this scenario in the subfamily of Lozi maps
    ($\tau_L = -\tau_R$, $\delta_L = \delta_R$) for his novel rigorous demonstration
    of robust chaos.
    \item With $\delta_L>0$ and $\delta_R<0$ \eqref{eq:BCNF2} is non-invertible
    mapping both the left half-plane ($x \le 0$)
    and the right half-plane ($x \ge 0$)
    onto the upper half-plane ($y \ge 0$).
    This scenario applies to border-collision bifurcations in a wide range of power
    converters whereby a switch in a circuit creates non-invertible dynamics
    \cite{BaVe01,DiTs02,ZhMo03}.
    \item With $\delta_L<0$ and $\delta_R>0$ \eqref{eq:BCNF2} is again non-invertible
    with applications to power electronics, but now maps to the lower half-plane ($y \le 0$)
    and exhibits different dynamical features (notice \eqref{eq:BCNF2}
    maps the origin to the right half-plane so does not have left/right symmetry).
\end{itemize}

Despite its apparent simplicity \eqref{eq:BCNF2}
has an incredibly complex bifurcation structure
that remains to be fully understood \cite{BaGr98,FaSi23,SiMe08b,SuGa08,ZhMo06b}.
The most significant bifurcations are those at which the attractor splits into pieces.
In~\cite{GhSi22} we used renormalisation
to uncover an infinite sequence of codimension-one bifurcations.
However, this only accommodated the orientating-preserving
setting of Banerjee {\em et al}~\cite{BaYo98}.
In this paper we identify similar sequences of bifurcations
in the orientation-reversing and non-invertible parameter regimes.
Again we use renormalisation to find bifurcations,
but now this approach misses some bifurcations.
For this reason we combine the analytical framework
with brute-force numerical simulations that
compute the number of connected components
from the behaviour of forward orbits.

The remainder of this paper is organised as follows.
We start in \S\ref{sec:1d} by reviewing the bifurcation structure
of one-dimensional piecewise-linear maps (skew tent maps)
and how this structure can be generated through renormalisation.
This corresponds to the BCNF in the special case $\delta_L = \delta_R = 0$,
and it is helpful to view the bifurcation structures described
in later sections as
extensions or perturbations from the structure that arises in one dimension.
Then in \S\ref{sec:bcnf} we review the parameter region $\Phi \subset \mathbb{R}^4$
where the BCNF has two saddle fixed points.
We describe special points on the stable and unstable manifolds of the fixed points
whose relation to one another can be used to characterise the occurrence
of homoclinic and heteroclinic bifurcations where the attractor is destroyed.
In \S\ref{sec:gen_renormalisation} we introduce the renormalisation scheme
and describe some basic aspects of this scheme that hold for all values of $\delta_L$ and $\delta_R$.
Then in \S\ref{sec:or_pr} we summarise the results of~\cite{GhSi22}
for the orientation-preserving setting.

Next in \S\ref{sec:numerics} we describe an algorithm
for numerically determining the number of connected components of attractor.
The algorithm outputs the greatest common divisor of a set of iteration numbers 
required for an orbit to return close to its starting point.
The algorithm is effective when the components of attractor are not too close to one another
and the dynamics on the attractor is ergodic.
Applied to the orientation-preserving setting
it reproduces the bifurcation structure obtained by renormalisation.

In \S\ref{sec:or_re}, \S\ref{sec:nonInvertNice}, and \S\ref{sec:nonInvertNotNice}
we study the orientation-reversing and non-invertible cases
and explain why the same renormalisation scheme should be
expected to work in these settings.
Formal proofs are beyond the scope of this study because,
as evident from~\cite{GhSi22},
these require a detailed analysis of the four-dimensional nonlinear
renormalisation operator,
plus will involve additional complexities because
now stable period-two and period-four solutions are possible.
Also if $\delta_L < 0$ and $\delta_R > 0$ then
the bifurcations that destroy the attractor
are different and more difficult to characterise.
In \S\ref{sec:reduction} we describe a special case unique
to the non-invertible settings where the dynamics reduces to one dimension.
Concluding remarks are provided in \S\ref{sec:conc}.

\section{Skew tent maps}
\label{sec:1d}
\begin{figure}[b!]
\begin{tabular}{cc}
  \includegraphics[scale=0.5]{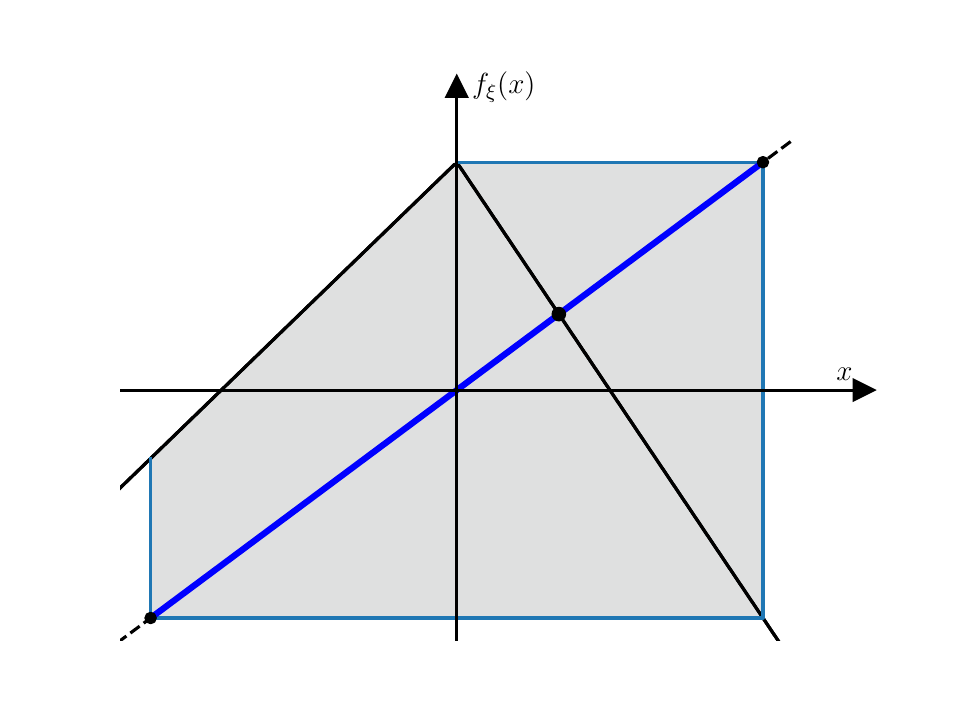} &   \includegraphics[scale=0.5]{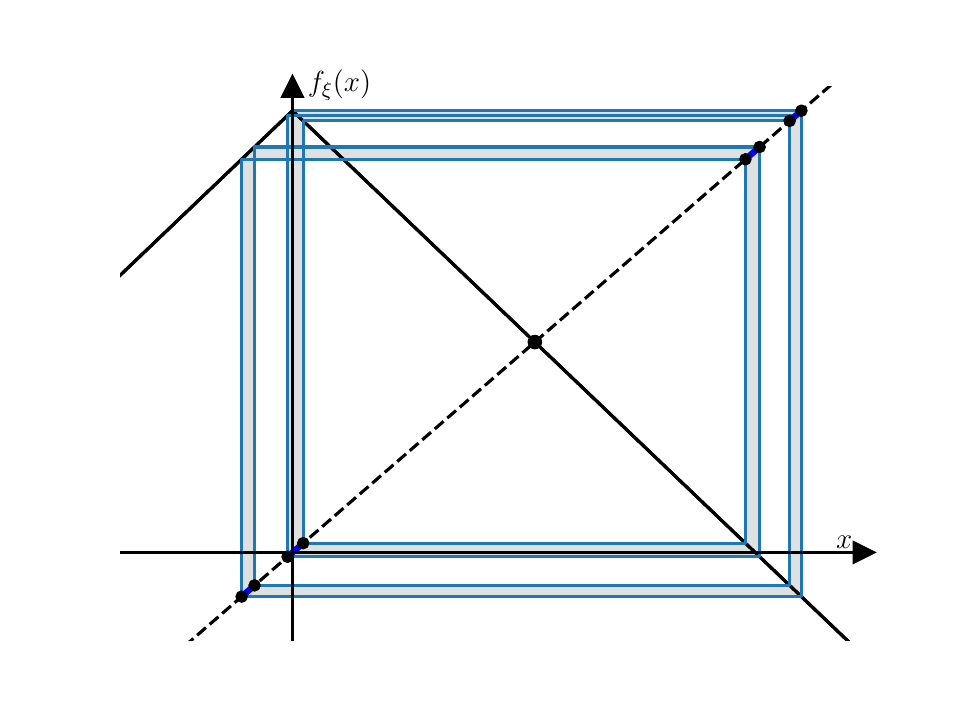}\\
(a) $(\tau_L,\tau_R) = (1.3,-2)$. &
(b) $(\tau_L,\tau_R) = (1.1,-1.1)$. \\[3pt]
\end{tabular}
\caption{Cobweb diagrams showing the attractor
of the skew tent map \eqref{eq:skewtent_map}
with two different combinations of the parameter values.
In (a) the attractor is an interval;
in (b) the attractor is the union of four disjoint intervals.
The maps are also instances of the BCNF
\eqref{eq:BCNF2} with $\xi = (1.3,0,-2,0)$ in panel (a)
and $\xi = (1.1,0,-1.1,0)$ in panel (b).
}
\label{fig:cobwebs_skewtent}
\end{figure}

With $\delta_L = \delta_R = 0$ the $y$-dynamics of~\eqref{eq:BCNF2} is trivial
and the $x$-dynamics reduces to
\begin{align}
\label{eq:skewtent_map}
    x \mapsto \begin{cases} \tau_L x + 1, & x \le 0, \\ \tau_R x + 1, & x \ge 0. \end{cases}
\end{align}
This is a two-parameter family of skew tent maps.
For any values of the parameters $\tau_L$ and $\tau_R$
for which \eqref{eq:skewtent_map} has an attractor,
this attractor is unique.
Fig.~\ref{fig:cobwebs_skewtent} shows typical examples
of the attractor using $\tau_L > 1$ and $\tau_R < -1$
with which \eqref{eq:skewtent_map} is piecewise-expanding,
so the attractor is chaotic
in a strict sense by the results of Lasota and Yorke \cite{LaYo73}
and Li and Yorke \cite{LiYo78}.
Fig.~\ref{fig:cobwebs_skewtent} shows typical examples of the chaotic attractor.
In panel (a) the attractor is
the interval $[\tau_R+1,1]$ whose endpoints are
the first and second iterates of the switching value $x=0$.
In panel (b) the attractor is a disjoint union of four intervals
whose endpoints are the first through eighth iterates of $x=0$.

\begin{figure}[b!]
\vskip 6pt
\centering
\includegraphics[width=0.8\linewidth]{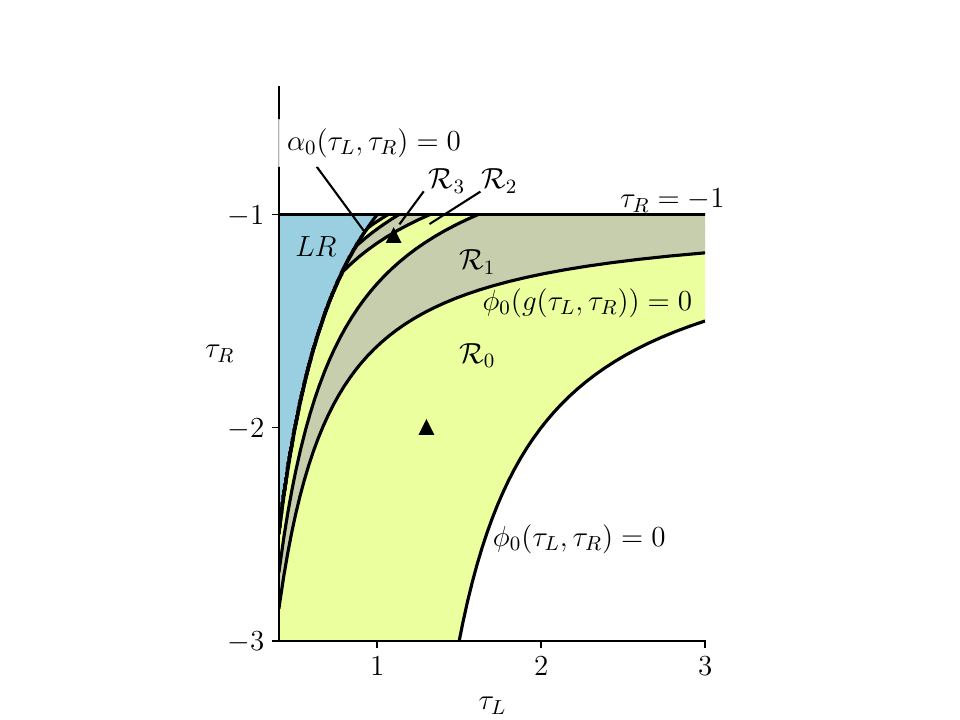}
\caption{A two-parameter bifurcation diagram of the skew tent map family~\eqref{eq:skewtent_map}.
In $\mathcal{R}_0$ the attractor is an interval.
In each $\mathcal{R}_n$ with $n \ge 1$ the attractor is
comprised of $2^n$ disjoint intervals. Below $\phi_0(\tau_L,\tau_R) = 0$ the map has no attractor; above $\tau_R = -1$ it has a stable fixed point in $x>0$; in the top-left region it has a stable $LR$-cycle (period-two solution).
The triangles indicate the parameter values used in Fig.~\ref{fig:cobwebs_skewtent}.
}\label{fig:skewtent_parameterspace}
\end{figure}

For all values of $\tau_L$ and $\tau_R$
the nature of the attractor is well understood~\cite{MaMa93,NuYo95,SuAv16}. Fig.~\ref{fig:skewtent_parameterspace} shows how
the part of the $(\tau_L,\tau_R)$-plane that is
relevent to us divides into regions according to the number of intervals
that comprise the attractor. In the top-left region labelled $LR$
the map has a stable period-two solution.
This solution has one point in the left half-plane
and one point in the right half-plane, so is termed an $LR$-cycle.
This region is bounded by the curve $\alpha_0(\tau_L,\tau_R) = 0$, where
\begin{align}
    \alpha_0(\tau_L, \tau_R) = \tau_L\tau_R+1,
    \label{eq:alpha0}
\end{align}
which is where the $LR$-cycle loses stability
by attaining a stability multiplier of $-1$.
The boundary in the lower-right part of the figure
is a homoclinic bifurcation where $x=0$ maps in two iterations to the fixed point of \eqref{eq:skewtent_map} in $x < 0$.  This occurs on the curve $\phi_0(\tau_L,\tau_R) = 0$, where
\begin{align}
   \phi_0(\tau_L,\tau_R) = \tau_L \tau_R + \tau_L - \tau_R.
\end{align}

As shown by Ito {\em et al} \cite{ItTa79,ItTa79b} the remaining boundaries can be identified through renormalisation as follows. Suppose there exists an interval neighbourhood $I$ of $x=0$ that maps into $x > 0$, then, on the second iteration, back into $I$.
In this case the restriction of the second iterate of \eqref{eq:skewtent_map} to $I$ is piecewise-linear with two pieces.  The $x<0$ piece corresponds to iterates with symbolic itinerary $LR$ and has slope $\tau_L \tau_R < -1$, while the $x>0$ piece corresponds to iterates with symbolic itinerary $RR$ and has slope $\tau_R^2 > 1$.  Thus the second iterate of~\eqref{eq:skewtent_map} on $I$ is conjugate to~\eqref{eq:skewtent_map} with $\tau_R^2$ in place of $\tau_L$ and $\tau_L \tau_R$ in place of $\tau_R$.  This corresponds to the {\em substitution rule} $(L,R) \mapsto (RR,LR)$, and induces the {\em renormalisation operator}
\begin{align}
\label{eq:skewtent_renorm}
g(\tau_L,\tau_R) = (\tau_R^2,\tau_L \tau_R).
\end{align}
By using the preimages of the
homoclinic bifurcation boundary we define
the sequence of regions
    \begin{align}
    \cR_n = \left\{ (\tau_L,\tau_R) \,\middle|\, \tau_R < -1,
    \phi_0 \left( g^n(\xi) \right) > 0, \phi_0 \left( g^{n+1}(\xi) \right) \le 0, \alpha_0(\tau_L,\tau_R) < 0 \right\},
    \label{eq:RnskewTentMap}
\end{align}
shown in Fig.~\ref{fig:skewtent_parameterspace}.
These are non-empty for all $n \ge 0$
and converge to $(\tau_L,\tau_R) = (1,-1)$ as $n \to \infty$.
In $\cR_0$ the attractor consists of one interval:

\begin{theorem}
    For any $(\tau_L,\tau_R) \in \cR_0$ with $\tau_L \ge 1$,
    the interval $[\tau_R+1,1]$ is the unique attractor.
    \label{th:R0skewTentMap}
\end{theorem}

A simple proof of Theorem \ref{th:R0skewTentMap} can be found
in Veitch and Glendinning \cite{VeGl90}.
The condition $\tau_L \ge 1$ is needed because for some $0 < \tau_L < 1$
and $\tau_R < -1$ the attractor is periodic.
This condition can be weakened but this is not needed for our purposes.

To characterise the attractor in the other regions $\cR_n$
we first make the following observation
that follows simply from
the definitions of $g$ and $\cR_n$.

\begin{proposition}
    If $(\tau_L,\tau_R) \in \cR_n$ with $n \ge 1$,
    then $g(\tau_L,\tau_R) \in \cR_{n-1}$.
\end{proposition}

Thus if $(\tau_L,\tau_R) \in \cR_n$
then $g^n(\tau_L,\tau_R) \in \cR_0$.
Theorem \ref{th:R0skewTentMap} can be applied
at this parameter point because any point below $\tau_R = -1$
maps under $g$ to the right of $\tau_L = 1$.
Moreover, the above conjugacy can be shown to hold
under all $n$ applications of $g$, see Ito {\em et al} \cite{ItTa79,ItTa79b},
thus the $2^n$-iterate of \eqref{eq:skewtent_map}
has an interval attractor.
The images of this interval under \eqref{eq:skewtent_map}
give $2^n$ disjoint intervals and hence the following result.

\begin{theorem}
\label{thm:skewtent_renorm}
If $(\tau_L,\tau_R) \in \cR_n$ with $n \ge 1$,
then \eqref{eq:skewtent_map} has a unique attractor comprised of $2^n$ disjoint intervals.
\end{theorem}

\section{Two-dimensional piecewise-linear maps with two saddle fixed points}
\label{sec:bcnf}

We now return to the BCNF \eqref{eq:BCNF2} with non-zero values of $\delta_L$ and $\delta_R$.  
In this section we identify two saddle fixed points and
compute some important points on their stable and unstable manifolds.

Motivated by Banerjee {\em et al} \cite{BaYo98} we focus on the following subset of four-dimensional parameter space:
\begin{align}
\label{eq:Phi}
    \Phi = \left\{\xi \in \mathbb{R}^4\ \middle|\ \tau_L > |\delta_L + 1|, \tau_R < -|\delta_R + 1| \right\}.
\end{align}
It is a simple exercise to show that $\Phi$ is the set of all parameter values for which~\eqref{eq:BCNF2} has two saddle fixed points~\cite{GhMc23}. These fixed points are
\begin{align}
X &= \left(\frac{-1}{\tau_R-\delta_R-1}, \frac{\delta_R}{\tau_R-\delta_R-1}\right),
\label{eq:X} \\
Y &= \left(\frac{-1}{\tau_L-\delta_L-1}, \frac{\delta_L}{\tau_L-\delta_L-1}\right),
\label{eq:Y}
\end{align}
where $X$ is in the open right half-plane $x > 0$,
and $Y$ is in the open left half-plane $x < 0$. Let 
\begin{align}
A_L &= \begin{bmatrix} \tau_L & 1 \\ -\delta_L & 0 \end{bmatrix}, &
A_R &= \begin{bmatrix} \tau_R & 1 \\ -\delta_R & 0 \end{bmatrix},
\label{eq:ALAR}
\end{align}
denote the Jacobian matrices associated with $Y$ and $X$ respectively.
With $\xi \in \Phi$ the matrix $A_L$ has eigenvalues $|\lambda_L^s|<1$ and $\lambda_L^u > 1$, while $A_R$ has eigenvalues $|\lambda_R^s| < 1$ and $\lambda_R^u < -1$.

\begin{figure}
\begin{center}
\begin{tabular}{cc}
  \includegraphics[scale=0.48]{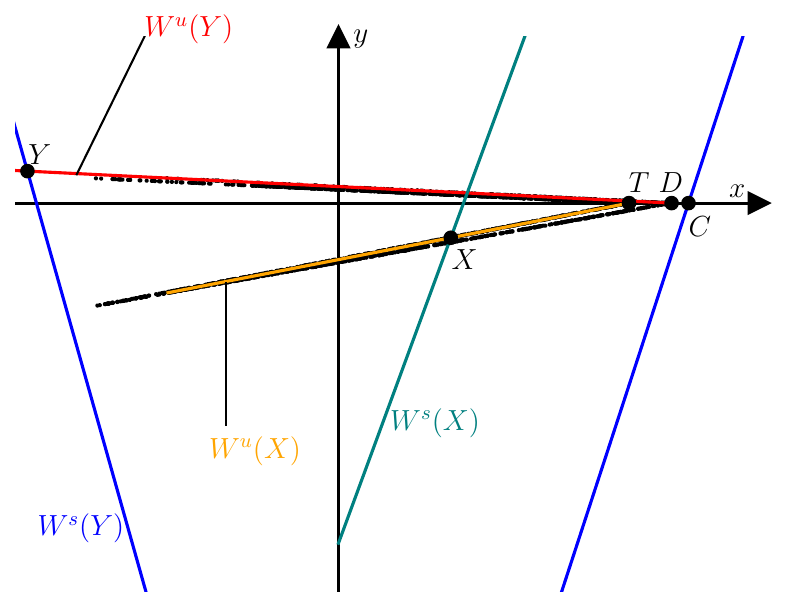} &  \includegraphics[scale=0.48]{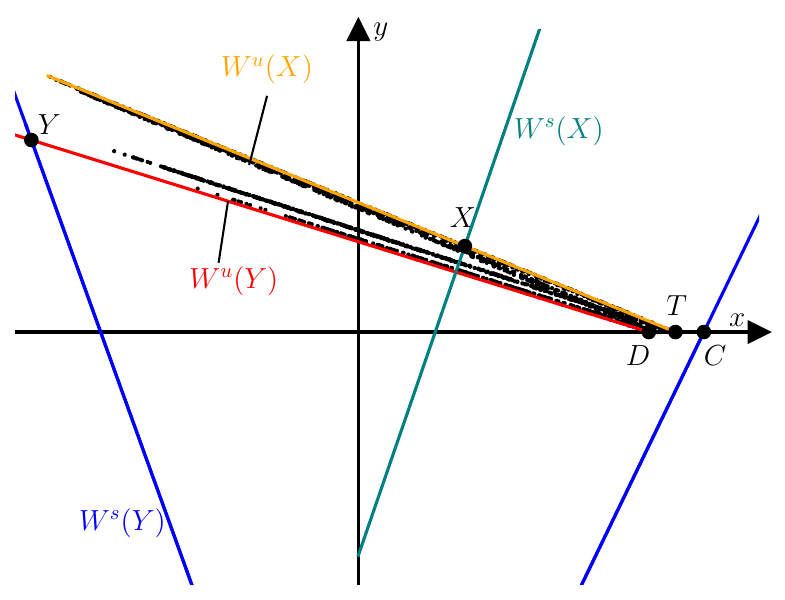} \\
(a) $\delta_L>0, \delta_R>0$ & (b) $\delta_L>0, \delta_R<0$ \\[3pt]
 \includegraphics[scale=0.48]{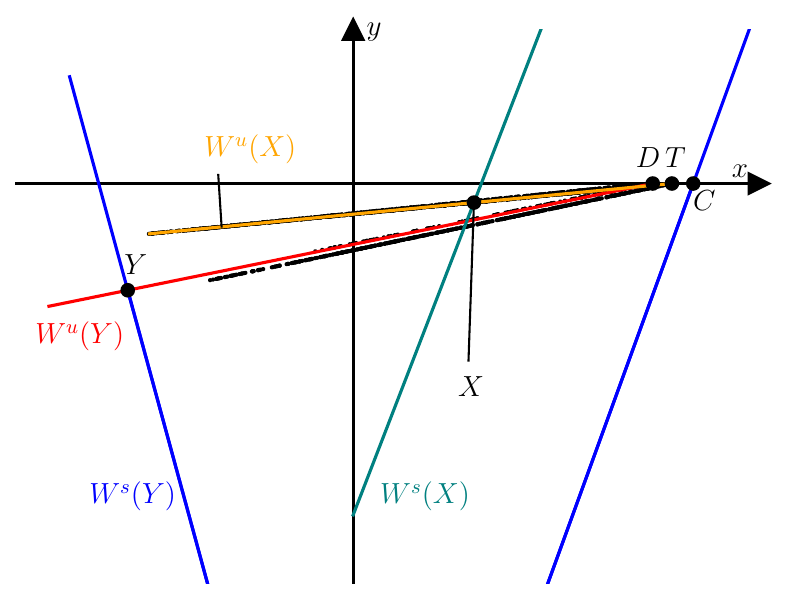} &   \includegraphics[scale=0.48]{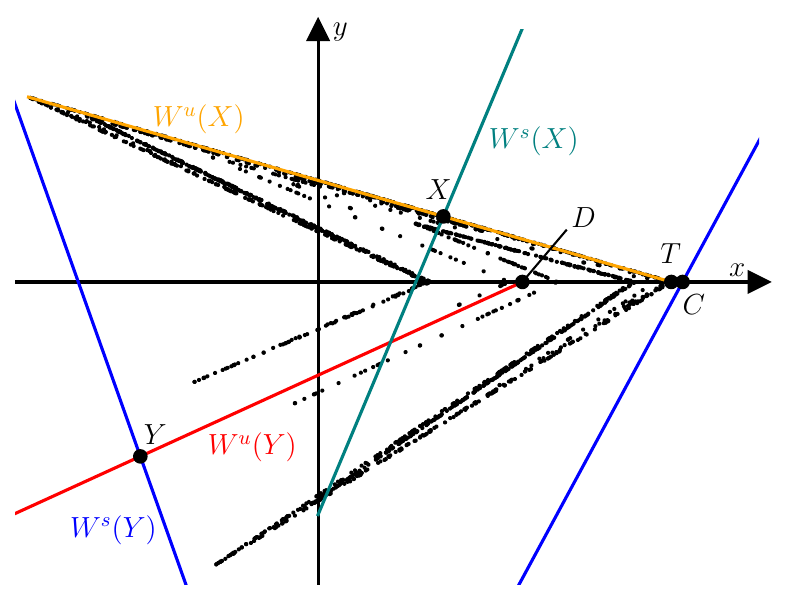} \\
(c) $\delta_L<0, \delta_R>0$ & (d) $\delta_L<0, \delta_R<0$ \\[3pt]
\end{tabular}
\end{center}
\caption{
Phase portraits of the BCNF \eqref{eq:BCNF2} for four different parameter
combinations: (a) $\xi = (2.1, 0.06, -1.7, 0.18)$; (b) $\xi = (2.1, 0.4, -1.7, -0.55)$; (c) $\xi = (2.2, -0.3, -1.7, 0.1)$; (d) $\xi = (1.8, -0.75, -1.6, -0.4)$.
Each plot shows the linear segments of the stable and unstable manifolds emanating from
the fixed points $X$ and $Y$, as well as an
adjoining segment of the stable manifold
of $Y$. We also indicate some of their intersections with $y=0$:
$T$, $D$, and $C$; formulas for these are given by
\eqref{eq:T}, \eqref{eq:D}, and \eqref{eq:C}.
To illustrate the chaotic attractor the black dots show
2000 iterates of a
typical forward orbit after transient dynamics has decayed.
\label{fig:pp}
}
\end{figure}

Since $X$ and $Y$ are saddles their stable and unstable manifolds are one-dimensional, and since the map is piecewise-linear these manifolds are piecewise-linear. Their stable manifolds $W^s(X)$ and $W^s(Y)$ have kinks on the switching manifold $x=0$ and at preimages of these points, while their unstable manifolds $W^u(X)$ and $W^u(Y)$ have kinks on the image of the switching manifold $y=0$ and at images of these points. For instance, as we grow $W^u(X)$ outwards from $X$, in one direction its first kink occurs at the point
\begin{align}
\label{eq:T}
T = \left(\frac{1}{1-\lambda_R^s}, 0 \right),
\end{align}
on $y=0$, see Fig.~\ref{fig:pp}.
Similarly as we grow $W^u(Y)$ outwards from $Y$, in one direction its first kink occurs at
\begin{align}
\label{eq:D}
D = \left(\frac{1}{1 - \lambda_L^s}, 0 \right).
\end{align}
A third point $C$ will be also central to our later calculations.  This point is where $W^s(Y)$, when grown outwards from $Y$ to the right, first intersects $y=0$ at a point with $x > 0$ and is given by
\begin{align}
\label{eq:C}
C = \left( \frac{{\lambda_L^u}^2}{(\delta_R - \tau_R \lambda_L^u)(\lambda_L^u - 1)}, 0\right).
\end{align}
Fig.~\ref{fig:pp} indicates the points $T$, $D$, and $C$
for four example phase portraits,
one for each of the four cases for the signs of $\delta_L$ and $\delta_R$.
Notice if $\delta_L > 0$ (upper plots)
the left half-plane maps to the upper half-plane,
while if $\delta_L < 0$ (lower plots)
the left half-plane maps to the lower half-plane.
Similarly if $\delta_R > 0$ (left plots)
the right half-plane maps to the lower half-plane,
while if $\delta_R < 0$ (right plots)
the right half-plane maps to the upper half-plane.

\section{Renormalisation in two dimensions}
\label{sec:gen_renormalisation}

In \S\ref{sec:1d} we saw that renormalisation based on the substitution rule $(L, R) \mapsto (RR, LR)$ explains the bifurcation structure of the
one-dimensional skew tent map family shown in
Fig.~\ref{fig:skewtent_parameterspace}.
For the BCNF \eqref{eq:BCNF2} this rule defines the renormalisation operator 
\begin{align}
\label{eq:g_xi}
g(\xi) = \left(\tau_R^2 - 2\delta_R, \delta_R^2, \tau_L\tau_R - \delta_L -\delta_R, \delta_L\delta_R \right).
\end{align}
The four values on the right-hand side of \eqref{eq:g_xi} are the traces and determinants of $A_R^2$ and $A_RA_L$. In this section we extend the basic aspects of this renormalisation operator beyond the orientation-preserving setting
of our earlier work \cite{GhSi22}. Renormalisation has also been applied to subfamilies of~\eqref{eq:BCNF2} by Pumari\~{n}o {\em et al} \cite{PuRo18, PuRo19} and Ou~\cite{Ou22}. 

We first identify the subset of $\Phi$ where the BCNF
has a stable $LR$-cycle.
Generalising \eqref{eq:alpha0} we have that the matrix $A_L A_R$
(whose eigenvalues are the stability multipliers of the $LR$-cycle)
has an eigenvalue $-1$ when $\alpha(\xi) = 0$, where
\begin{align}
\label{eq:gen_alpha}
\alpha(\xi) = \tau_L\tau_R + (\delta_L-1)(\delta_R-1).
\end{align}
If $\alpha(\xi) > 0$ this eigenvalue is greater than $-1$.
For the $LR$-cycle to be stable we also need $\det(A_L A_R) < 1$,
where $\det(A_L A_R) = \delta_L \delta_R$, so we define
\begin{equation}
    \mathcal{P}_2 = \left\{ \xi \in \Phi \,\middle|\, \delta_L\delta_R<1,\, \alpha(\xi)>0 \right\}.
\end{equation}

\begin{proposition}
\label{pr:g_doesnt_map}
If $\xi \in \mathcal{P}_2$ then $f_\xi$ has an asymptotically stable $LR$-cycle.
\end{proposition}

\begin{proof}
Let $\tilde{\tau}_R = \tau_L \tau_R - \delta_L - \delta_R$
and $\tilde{\delta}_R = \delta_L \delta_R$
be the trace and determinant of $A_L A_R$.
All eigenvalues of $A_L A_R$ have modulus less than $1$
if and only if $\tilde{\delta}_R < 1$ and $-\tilde{\delta}_R-1 < \tilde{\tau}_R < \tilde{\delta}_R+1$.
Certainly $\tilde{\delta}_R<1$ by assumption, and
$-\tilde{\delta}_R-1 < \tilde{\tau}_R$ by the definition of $\alpha$;
also
$$
\tilde{\tau}_R - \tilde{\delta}_R - 1
= \tau_L\tau_R - (\delta_L+1)(\delta_R+1)
< -|(\delta_L+1)(\delta_R+1)| - (\delta_L+1)(\delta_R+1)
\le 0
$$
by the definition of $\Phi$.

The composed map $f_R \circ f_L$
is affine with unique fixed point
$$
P = \frac{-1}{\tilde{\tau}_R - \tilde{\delta}_R - 1} \big( \tau_R+\delta_R+1, -\delta_R(\tau_L+\delta_L+1) \big).
$$
Let
$$
Q = f_L(P) = \frac{-1}{\tilde{\tau}_R - \tilde{\delta}_R - 1} \big( \tau_L+\delta_L+1, -\delta_L(\tau_R+\delta_R+1) \big).
$$
Notice the first component of $P$ is negative
while the first component of $Q$ is positive,
thus $\{ P, Q \}$ is a periodic solution of $f_\xi$.
Above we showed all eigenvalues of $A_L A_R$
have modulus less than $1$, hence $\{ P, Q \}$ is asymptotically stable.
\end{proof}

If $\alpha(\xi) < 0$ the $LR$-cycle is unstable.
This is always the case if $\delta_L + \delta_R \ge 0$:

\begin{proposition}
\label{pr:0}
If $\xi \in \Phi$ with $\delta_L + \delta_R \ge 0$ then $\alpha(\xi)<0$.
\end{proposition}

\begin{proof}
Suppose $\xi \in \Phi$ with $\delta_L + \delta_R \ge 0$.
If $\delta_L, \delta_R \ge -1$ then
$$
\alpha(\xi) < -(\delta_L+1)(\delta_R+1) + (\delta_L-1)(\delta_R-1)
= -2 (\delta_L + \delta_R) \le 0,
$$
using the definition of $\Phi$ in the first inequality.
If instead $\delta_L < -1$ or $\delta_R < -1$
then $\delta_L \delta_R < -1$ because $\delta_L + \delta_R > 0$, so
$$
\alpha(\xi) < (\delta_L+1)(\delta_R+1) + (\delta_L-1)(\delta_R-1)
= 2 (\delta_L \delta_R + 1) < 0.
$$
\end{proof}

Now we show that if $\alpha(\xi) < 0$
then the renormalisation operator $g$
produces another instance of the BCNF in $\Phi$.

\begin{proposition}
\label{pr:gmaps}
If $\xi \in \Phi$ and $\alpha(\xi)<0$ then $g(\xi) \in \Phi$.
\end{proposition}

\begin{proof}
Let $\tilde{\tau}_L = \tau_R^2 - 2 \delta_R$,
let $\tilde{\delta}_L = \delta_R^2$,
and let $\tilde{\tau}_R$ and $\tilde{\delta}_R$ be as in the proof
of Proposition \ref{pr:g_doesnt_map};
then $g(\xi) = \big( \tilde{\tau}_L, \tilde{\delta}_L, \tilde{\tau}_R, \tilde{\delta}_R \big)$.
Observe $\tilde{\tau}_L - \tilde{\delta}_L - 1 = \tau_R^2 - (\delta_R+1)^2 > 0$ because $\tau_R < -(\delta_R + 1)$,
and $\tilde{\tau}_L + \tilde{\delta}_L + 1 = \tau_R^2 + (\delta_R+1)^2 > 0$ because $\tau_R < 0$, thus $\tilde{\tau}_L > |\tilde{\delta}_L + 1|$.
Also $\tilde{\tau}_R - \tilde{\delta}_R - 1 = \tau_L \tau_R - (\delta_L+1)(\delta_R+1) < 0$ because $\tau_L > \delta_L + 1$ and $\tau_R < \delta_R + 1$, and $\tilde{\tau}_R + \tilde{\delta}_R + 1 = \alpha(\xi) < 0$, by assumption, thus $\tilde{\tau}_R < -|\tilde{\delta}_R + 1|$.
\end{proof}

\begin{figure}[h]
\begin{tabular}{cc}
  \includegraphics[scale=0.5]{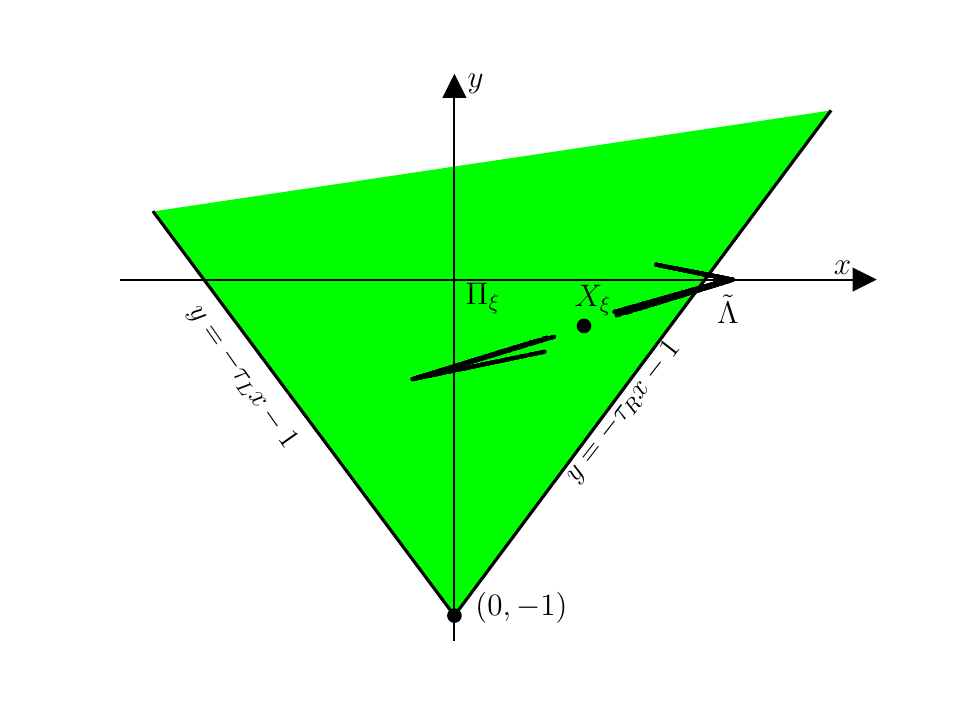} &   \includegraphics[scale=0.5]{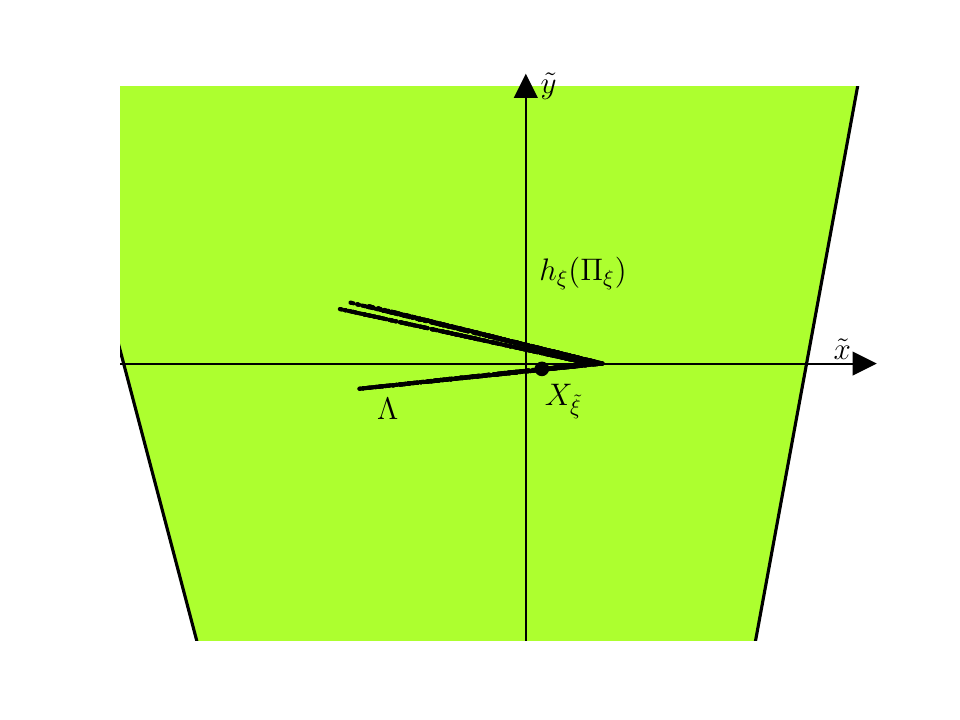}\\
(a) $\xi=\xi^{(1)}_{\rm ex} \in \cR^{(1)}_1$ & (b) $\xi = g \big( \xi^{(1)}_{\rm ex} \big) \in \cR^{(1)}_0$ \\[3pt]
\end{tabular}
\caption{
Phase portraits of the BCNF \eqref{eq:BCNF2}.
Panel (a) uses the example parameter point
$\xi^{(1)}_{\rm ex}$ given by \eqref{eq:xi1ex}
where the attracor $\tilde{\Lambda}$ has two connected
components, one of which lies entirely in $\Pi_\xi$ (shaded).
Panel (b) uses the parameter point $g \big( \xi^{(1)}_{\rm ex} \big)$
where the attractor $\Lambda$ has one connected component
in $h_\xi(\Pi_\xi)$ (shaded).
}
\label{fig:Pi}
\end{figure}

Now let
\begin{equation}
\Pi_\xi = \left\{ f_\xi^{-1}(x,y) \,\middle|\, x \ge 0 \right\},
\label{eq:gen_Pi}
\end{equation}
be the set of all points that map to the right half-plane.
On $\Pi_\xi$ the second iterate $f_\xi^2$ has only two pieces:
\begin{equation}
f_\xi^2(x,y) =
\begin{cases}
\left( f_{R,\xi} \circ f_{L,\xi} \right)(x,y), & x \le 0, \\
f_{R,\xi}^2(x,y), & x \ge 0.
\end{cases}
\label{eq:gen_BCNF22}
\end{equation}
For any $\xi \in \Phi$ \eqref{eq:gen_BCNF22} is affinely conjugate to
$f_{g(\xi)}$. Specifically
$f_\xi^2 = h_\xi^{-1} \circ f_{g(\xi)} \circ h_\xi$ on $\Pi_\xi$,
where
\begin{equation}
h_\xi(x,y) = \frac{1}{\tau_R + \delta_R + 1} \begin{bmatrix} x \\ \delta_R x + \tau_R y - \delta_R \end{bmatrix},
\label{eq:h}
\end{equation}
is the necessary change of coordinates.
As an example Fig.~\ref{fig:Pi}-a
shows $\Pi_\xi$ (in the phase space of $f_\xi$)
at the parameter point
\begin{equation}
    \xi^{(1)}_{\rm ex} = (1.5, 0.4, -1.5, 0.4),
    \label{eq:xi1ex}
\end{equation}
while Fig.~\ref{fig:Pi}-b shows $h_\xi(\Pi_\xi)$
(in the phase space of $f_{g(\xi)}$).
The map $f_{g(\xi)}$ has an invariant set
$\Lambda \subset h_\xi(\Pi_\xi)$,
so, as shown below, $h_\xi^{-1}(\Lambda) \subset \Pi_\xi$
is an invariant set of $f_\xi^2$.
Moreover, the union of $h_\xi^{-1}(\Lambda)$
and its image under $f_\xi$ is an invariant set of $f_\xi$.
It is easy to infer the number of connected components in this set
from the number of connected components of $\Lambda$,
and this forms the basis of the renormalisation scheme.

\begin{proposition}
\label{pr:C}
Suppose $\Lambda \subset h_\xi(\Pi_\xi)$ is an invariant set of $f_{g(\xi)}$.
Then $\tilde{\Lambda} = h_\xi^{-1}(\Lambda) \cup f_\xi \left( h_\xi^{-1}(\Lambda) \right)$
is an invariant set of $f_\xi$.
Moreover, if $h_\xi^{-1}(\Lambda) \cap f_\xi \left( h_\xi^{-1}(\Lambda) \right) = \varnothing$
then the number of connected components of $\tilde{\Lambda}$ is twice
the number of connected components of $\Lambda$.
\end{proposition}

\begin{proof}
Let $S = h_\xi^{-1}(\Lambda)$.
Since $S \subset \Pi_\xi$,
\begin{equation}
f_\xi^2(S) = h_\xi^{-1} \left( f_{g(\xi)} \left( h_\xi(S) \right) \right).
\label{eq:f2S}    
\end{equation}
But $h_\xi(S) = \Lambda$ and $f_{g(\xi)}(\Lambda) = \Lambda$,
so the right-hand side of \eqref{eq:f2S} is $h_\xi^{-1}(\Lambda) = S$.
Thus $S$ is invariant under $f_\xi^2$, so
$S \cup f_\xi(S)$ is invariant under $f_\xi$.
Since $h_\xi$ is invertible the number of components
of $S$ is the same as the number of components of $\Lambda$.
If $S \cap f_\xi(S) = \varnothing$ then $f_\xi(S)$ also has this many components (for otherwise $f_\xi^2(S) = S$ would not be possible because $f_\xi$ is continuous).
\end{proof}

\section{The orientation-preserving case}
\label{sec:or_pr}

\begin{figure}
\begin{subfigure}{.5\linewidth}
\centering
\includegraphics[scale=0.48]{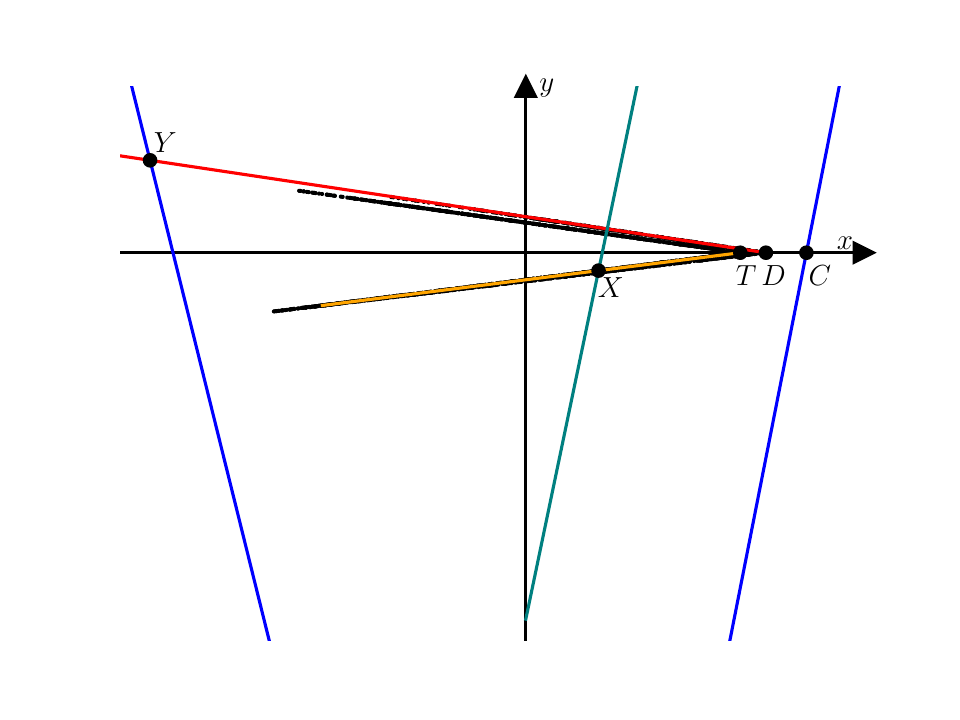}
\caption{$\tau_L = 1.7$}
\end{subfigure}%
\begin{subfigure}{.5\linewidth}
\centering
\includegraphics[scale=0.48]{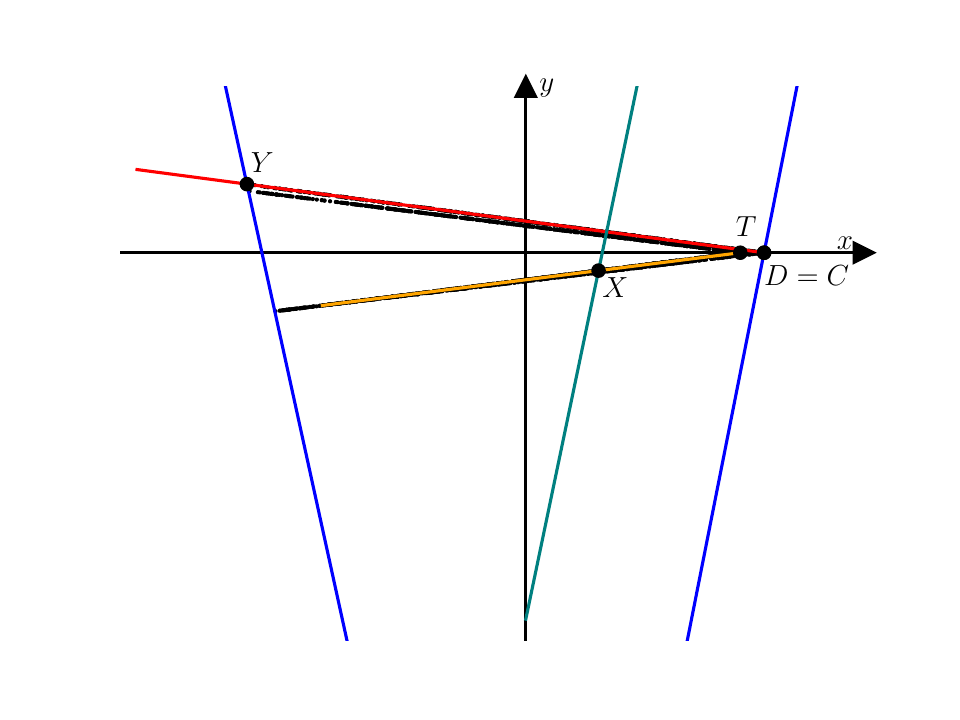}
\caption{$\tau_L \approx 1.9083$}
\end{subfigure}\\[1ex]
\begin{subfigure}{1\linewidth}
\centering
\includegraphics[scale=0.48]{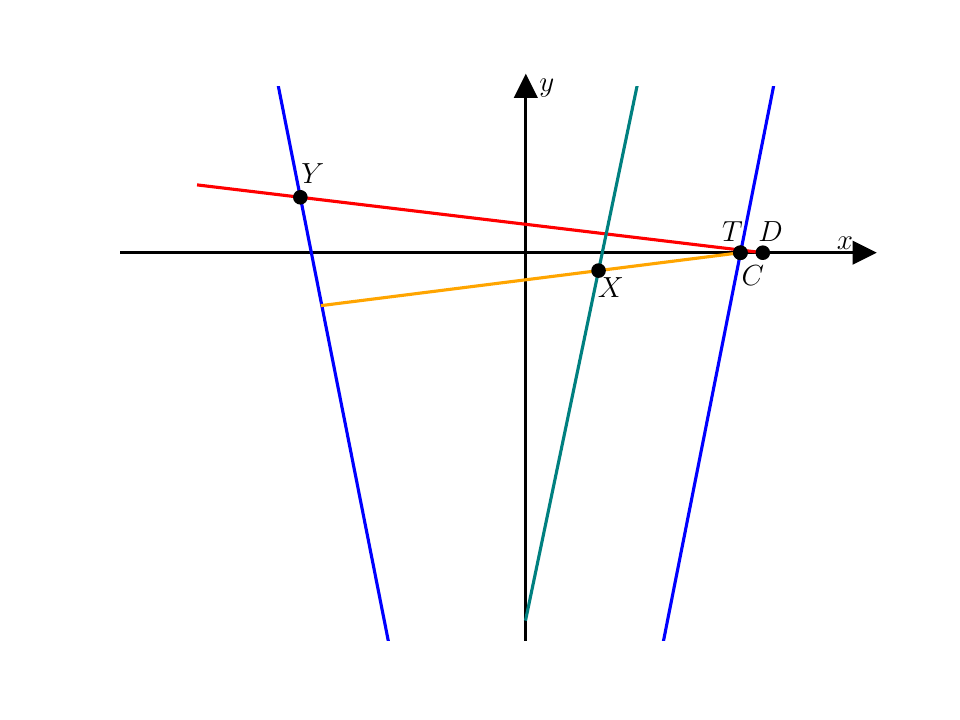}
\caption{$\tau_L = 2.1$}
\end{subfigure}
\caption{Phase portraits of \eqref{eq:BCNF2} with $\delta_L = 0.1$, $\delta_R=0.1$, $\tau_R=-2$ and three different values of $\tau_L$.
These parameter values correspond to the black triangles in Fig.~\ref{fig:reg1}-a.
Panel (b) uses $\tau_L$ such that $\phi^+(\xi) = 0$ to ten decimal places.
In (a) ${\rm cl}(W^u(X))$ is a chaotic attractor;
in (b) $W^s(Y)$ and $W^u(Y)$ form a homoclinic corner
by intersecting at $D=C$;
in (c) there is no attractor.}
\label{fig:typ_pp_1}
\end{figure}

In this section we summarise the main results of 
our earlier work \cite{GhSi22}.
Let
\begin{align}
\label{eq:Phi1}
    \Phi^{(1)} = \left\{\xi \in \Phi\ \middle|\ \delta_L >0, \delta_R > 0 \right\},
\end{align}
be the subset of $\Phi$ for which the BCNF is orientation-preserving.
In a large part of $\Phi^{(1)}$
an attractor is destroyed when the points $C$ and $D$ (shown in Fig.~\ref{fig:pp}-a)
coincide. This is a homoclinic bifurcation, or homoclinic corner \cite{Si16b},
where the stable and unstable manifolds of the fixed point $Y$
attain non-trivial intersections.
Straight-forward algebra gives
\begin{align}
\label{eq:CminusD}
C_1 - D_1 = \frac{\phi^+(\xi)}{(\lambda_L^u - 1)(1-\lambda_L^s)(\delta_R-\tau_R\lambda_L^u)},
\end{align}
where 
\begin{align}
\label{eq:phi+}
\phi^+(\xi)=\delta_R - (\tau_R + \delta_L + \delta_R - (1+\tau_R)\lambda_L^u)\lambda_L^u.
\end{align}
Fig.~\ref{fig:typ_pp_1} shows an example.
In panel (a) the closure of the unstable manifold of $X$,
\begin{align}
\label{eq:attractor}
    \Lambda = {\rm cl}(W^u(X)),
\end{align}
is a chaotic attractor.
As the value of $\tau_L$ is increased
the attractor approaches the point $D$ and is destroyed
at $\tau_L \approx 1.9083$
when $\phi^+(\xi) = 0$, panel (b).
At larger values of $\tau_L$
almost all forward orbits diverge, panel (c).


In some parts of $\Phi^{(1)}$ the attractor $\Lambda$
is not destroyed at $\phi^+(\xi) = 0$.
This occurs when it does not approach $D$
as $C \to D$; instead the attractor is destroyed in a subsequent
heteroclinic bifurcation involving a period-three solution \cite{GhSi22b}.

\begin{figure}[h]
\begin{tabular}{cc}
  \includegraphics[scale=0.5]{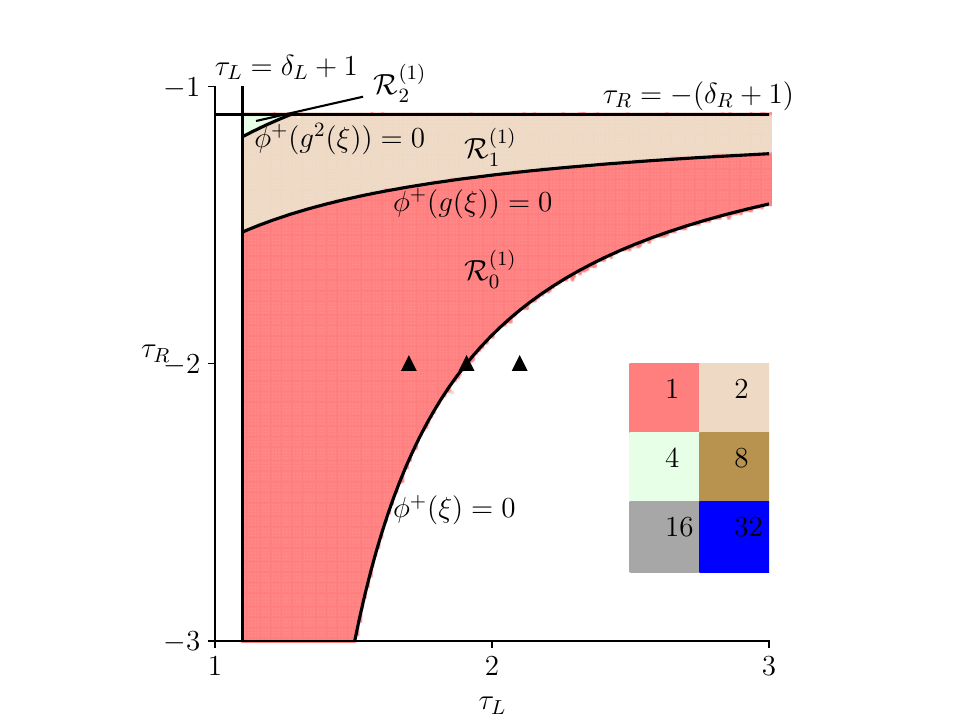} &   \includegraphics[scale=0.5]{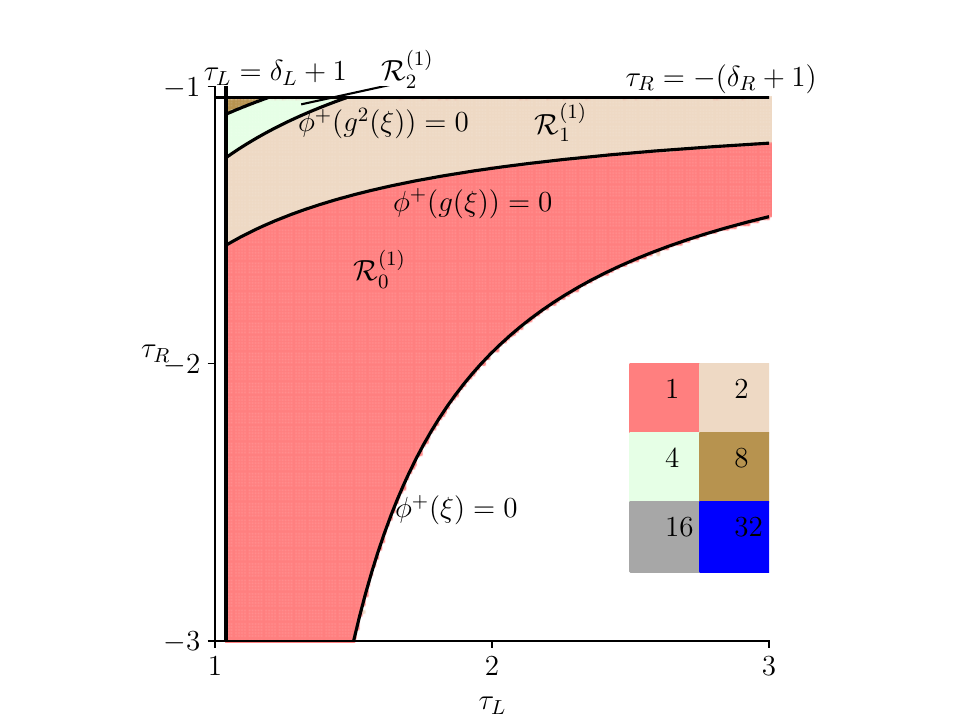}\\
(a) $\delta_L = 0.1, \delta_R=0.1$. & (b) $\delta_L = 0.04, \delta_R=0.04$.  \\[3pt]
\end{tabular}
\caption{
Two-dimensional slices of the
orientation-preserving parameter region $\Phi^{(1)}$
showing the curves $\phi(g^n(\xi)) = 0$, for low values of $n$,
overlaid upon the numerical results of Eckstein's greatest common divisor
algorithm, explained in \S\ref{sec:numerics}.
Each point in a $200 \times 200$ grid
is coloured by the greatest common divisor
of a set of close return iteration numbers $J$
according to the colour bar on the right.
In the algorithm we used $\varepsilon = 0.001$ and $M = 10^6$; also points are coloured white if iterates appeared to diverge.
The black triangles correspond to the parameter values of Fig.~\ref{fig:typ_pp_1}.
}
\label{fig:reg1}
\end{figure}

With $\delta_L = \delta_R = 0$
we have $\phi^+(\xi) = \tau_L \phi_0(\tau_L,\tau_R)$,
so in this case the bifurcation surface $\phi^+(\xi) = 0$
reduces to the homoclinic bifurcation
$\phi_0(\tau_L,\tau_R) = 0$ of the skew tent map family.
This motivates the definition
\begin{align}
\cR^{(1)}_n = \left\{\xi \in \Phi^{(1)} \,\middle|\, \phi^+(g^n(\xi)) > 0,\, \phi^+(g^{n+1}(\xi)) \le 0 \right\},
\end{align}
for all $n \ge 0$.
The constraint $\alpha(\xi) < 0$ is not needed in this definition
because, by Proposition \ref{pr:0}, if $\xi \in \Phi^{(1)}$
then automatically $\alpha(\xi) < 0$.
As shown in \cite{GhSi22} the regions $\cR^{(1)}_n$
are disjoint and cover the subset of $\Phi^{(1)}$
for which $\phi^+(\xi) > 0$.
These regions are more difficult to visualise than those in \S\ref{sec:1d}
because they are four-dimensional.
Fig.~\ref{fig:reg1} shows two-dimensional slices
of parameter space defined by fixing $\delta_L > 0$ and $\delta_R > 0$.
In these slices only finitely many $\cR^{(1)}_n$ are visible
because as $n \to \infty$ they converge to $\xi = (1,0,-1,0)$ (a fixed point of $g$).
The following result describes $\Lambda$
for parameter values in $\cR^{(1)}_0$.
For a proof see \cite{GhSi22}.

\begin{theorem}
For any $\xi \in \cR_0^{(1)}$,
\begin{enumerate}
\item[(i)]
$\Lambda$ is bounded, connected, and invariant under $f_\xi$,
\item[(ii)]
has a positive Lyapunov exponent, and
\item[(iii)]
if $\delta_R < 1$ there exists a forward invariant set $\Delta \subset \mathbb{R}^2$ with non-empty interior such that
\begin{equation}
\bigcap_{n=0}^\infty f_\xi^n(\Delta) = \Lambda.
\label{eq:gen_LambdaAsInfiniteIntersection}
\end{equation}
\end{enumerate}
\label{th:gen_R0}
\end{theorem}

Item (ii) says $\Lambda$ is chaotic in the sense
of having a positive Lyapunov exponent.
We believe it satisfies Devaney's definition of chaos
throughout $\Phi^{(1)}$, but have only proved this
in a subset of $\Phi^{(1)}$ \cite{GhSi22b}.
Item (iii) says $\Lambda$ is a Milnor attractor.
We believe the condition $\delta_R > 1$ is unnecessary
and $\Lambda$ is in fact a topological attractor
throughout $\cR^{(1)}_0$.

To understand the dynamics in all other regions $\cR_n^{(1)}$
we use the renormalisation operator $g$.
The following result follows simply
from Proposition \ref{pr:gmaps}
and the definition of $\cR^{(1)}_n$.

\begin{proposition}
\label{pr:gmaps1}
If $\xi \in \cR_n^{(1)}$ with $n \ge 1$, then $g(\xi) \in \cR_{n-1}^{(1)}$.
\end{proposition}

So if $\xi \in \cR_n^{(1)}$ then $g^n(\xi) \in \cR_0^{(1)}$
where the attractor is connected.
Then by $n$ applications of Proposition \ref{pr:C},
$f_\xi$ has an attractor with $2^n$ connected components
as long as the assumptions of Proposition \ref{pr:C} are
satisfied in each application.
This is indeed the case, as shown in \cite{GhSi22}
through a series of careful calculations, and gives the following result.

\begin{theorem}
For any $\xi \in \cR_n^{(1)}$ with $n \ge 0$,
$f_\xi$ has a chaotic Milnor attractor with exactly $2^n$ connected components.
\label{th:gen_affinelyConjugate}
\end{theorem}

For example Fig.~\ref{fig:Pi}-a uses
$\xi = \xi^{(1)}_{\rm ex}$ belonging to $\cR^{(1)}_1$
so the attractor has two connected components, as shown.
The renormalisation allows us to say more about this attractor.
Specifically each piece of the attractor is an affine transformation
of the one-component attractor of \eqref{eq:BCNF2} with
$\xi = g \big( \xi^{(1)}_{\rm ex} \big)$ belonging to $\cR^{(1)}_0$,
shown in Fig.~\ref{fig:Pi}-b.

\section{A component counting algorithm.}
\label{sec:numerics}

To support an extension of the theoretical results of \S\ref{sec:or_pr}
to the orientation-reversing and non-invertible settings,
we perform numerical simulations to
count the number of connected components of the attractors.
There are many methods for estimating the number of connected components
of a set $\Psi$ from a finite collection $F$ of points in $\Psi$.
For example Robins {\em et al} \cite{RoMe00}
connect all pairs of points in $F$ with line segments to form a complete graph,
then base their estimation from a minimal spanning tree.
In our setting $\Psi$ is generated by a map,
so it is more effective to use a method that utilises the dynamics.
For this reason we compute the number of components as the greatest common divisor
of a certain set of computed values.
This method originates with Eckstein \cite{Ec06}
and is described by Avrutin {\em et al} \cite{AvEc07}.
As explained at the end of this section,
the effectiveness of the method relies on the following result.

\begin{lemma}
\label{lem:att_components}
Suppose a compact invariant set $\Psi$ of a continuous map $f$ has $k \ge 2$ connected components, and $f$ has an orbit that visits all components.  Then the components can be labelled as $\Psi_1,\Psi_2,\ldots,\Psi_k$ such that $f(\Psi_1) = \Psi_2, f(\Psi_2) = \Psi_3, \ldots, f(\Psi_{k-1}) = \Psi_k$, and $f(\Psi_k) = \Psi_1$.
\end{lemma}

Thus the components of $\Psi$ are ordered cyclically, each has one `predecessor' and one `successor'.  The following proof is adapted from \cite{AvEc07,Ec06}.  It uses a small quantity $\varepsilon$ to avoid reference to path-connectedness.

\begin{proof}[Proof of Lemma.~\ref{lem:att_components}]
By assumption there exists $P \in \Psi$ whose forward orbit enters all components.  Let $\Psi_1$ be the component containing $P$.  Suppose for a contradiction $f(\Psi_1)$ contains points in two different components, $T$ and $U$.  Since $\Psi$ is compact there exists $d > 0$ such that any point in $T$ and any point in $U$ are at least a distance $d$ apart.  Since $\Psi_1$ is connected for any $\varepsilon > 0$ there exist $Q, R \in \Psi_1$ with $\| Q - R \| < \varepsilon$ such that $f(Q) \in T$ and $f(R) \in U$.  But $\| f(Q) - f(R) \| > d$ so this is not possible because $f$ is continuous, hence $f(\Psi_1)$ is a subset of one component of $\Psi$.  This component is not $\Psi_1$ (because then $\Psi_1$ would be forward invariant and the forward orbit of $P$ could not reach the other components), let us call it $\Psi_2$. So $f(\Psi_1) \subset \Psi_2$.

By a similar argument $f(\Psi_2)$ is a subset of one component, and if $k > 2$ this component is neither $\Psi_1$ nor $\Psi_2$, call it $\Psi_3$. Inductively we obtain $\Psi_1,\Psi_2,\ldots,\Psi_k$ with $f(\Psi_1) \subset \Psi_2, f(\Psi_2) \subset \Psi_3, \ldots, f(\Psi_{k-1}) \subset \Psi_k$.
Also $f(\Psi_k)$ is a subset of one component.

But $\Psi$ is invariant, meaning $f(\Psi) = \Psi$.  Thus $f(\Psi_k) \subset \Psi_1$, and further $f(\Psi_1) = \Psi_2, f(\Psi_2) = \Psi_3, \ldots, f(\Psi_{k-1}) = \Psi_k$, and $f(\Psi_k) = \Psi_1$ as required.
\end{proof}

Now fix $\xi$ and suppose $f_\xi$ has an attractor $\Lambda$
with $k \ge 1$ connected components.
To compute the value $k$, which is assumed to be unknown to us {\em a priori},
the algorithm proceeds as follows.
Fix $\ee > 0$ (we used $\ee = 0.001$ or $\ee=0.0001$)
and $M > 0$ (we used $M = 10^6$ )
and let $J = \varnothing$.
Choose some initial point assumed to be in the basin of attraction of $\Lambda$
and iterate it under $f_\xi$ a reasonably large number of times
(we used $10^4$ iterations) to remove transient dynamics
and obtain a point in $\Lambda$, or extremely close to $\Lambda$, call it $(x_0,y_0)$.
Iterate further, and for all $i = 1,2,\ldots,M$
evaluate the distance (Euclidean norm in $\mathbb{R}^2$)
between $f_\xi^i(x_0,y_0)$ and $(x_0,y_0)$.
If this distance is less than $\ee$, append the number $i$ to the set $J$.
Finally evaluate the greatest common divisor
of the elements in $J$ --- this is our estimate for the value of $k$.

For example using $\xi = \xi_{\rm ex}^{(1)}$, as in Fig.~\ref{fig:Pi}-a,
the algorithm generated a set of $1249$ numbers
$$
J = \{ 1292, 3170, 3778, \ldots, 999930 \},
$$
whose elements have greatest common divisor $2$
(and indeed at this parameter point the attractor has two components).

Fig.~\ref{fig:reg1} shows the output of this algorithm
over a $200 \times 200$ grid of parameter points.
The results show excellent agreement to the theory described in \S\ref{sec:or_pr}.
For example parameter points where the greatest common divisor
of the numbers in $J$ is two have boundaries indistinguishable from the true bifurcation boundaries
$\phi(g(\xi)) = 0$ and $\phi(g^2(\xi)) = 0$.

Two principles underlie the effectiveness of the algorithm.
First, if the distance between any two components of $\Lambda$ is greater than $\ee$,
then $\left\| f_\xi^i(x_0,y_0) - (x_0,y_0) \right\| < \ee$ implies that
$f_\xi^i(x_0,y_0)$ and $(x_0,y_0)$ belong to the same component of $\Lambda$
(assuming $(x_0,y_0) \in \Lambda$).
By Lemma \ref{lem:att_components} this is only possible if $i$ is a multiple of $k$.
Thus the elements of $J$ are all multiples of $k$.

Second, assuming $f_\xi$ is ergodic on $\Lambda$,
the set of all $i > 0$ giving $\left\| f_\xi^i(x_0,y_0) - (x_0,y_0) \right\| < \ee$
will not be expected to share a multiple larger than $k$.
This is because ergodicity implies the dynamics of $f_\xi^k$
on any component are well mixed
(see Haydn {\em et al} \cite{HaLa05} and references within for theory
on the probability distribution of the values of $i$ when $\ee$ is small).
In our setting $M = 10^6$ seems to generate large enough
sets $J$ to ensure
the greatest common divisor of the numbers in $J$ is $k$, instead a multiple of $k$.
One can also modify the algorithm, as described in \cite{AvEc07,Ec06},
to search for a close return to several reference points,
instead of the single point $(x_0,y_0)$.

\section{The orientation-reversing case}
\label{sec:or_re}

\begin{figure}
\begin{subfigure}{.5\linewidth}
\centering
\includegraphics[scale=0.48]{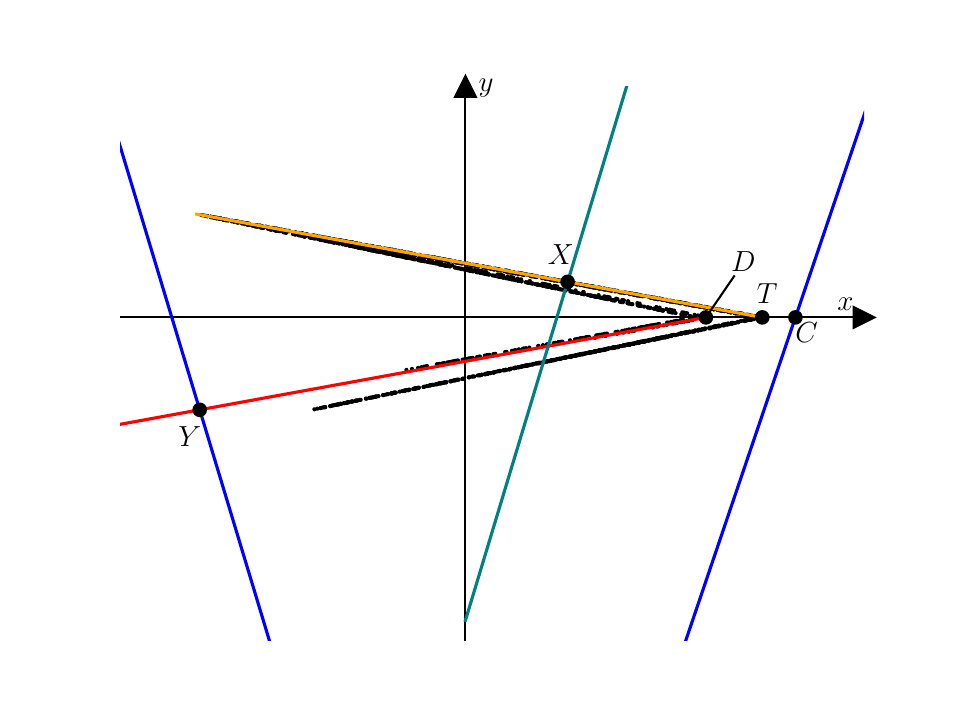}
\caption{$\tau_L = 1.8$}
\end{subfigure}%
\begin{subfigure}{.5\linewidth}
\centering
\includegraphics[scale=0.48]{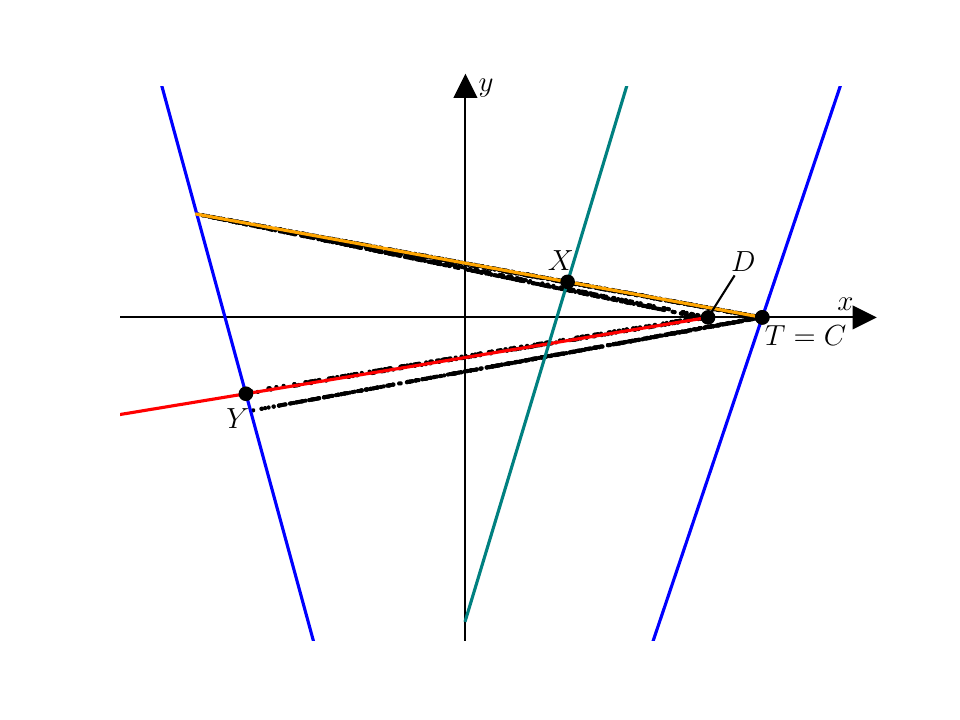}
\caption{$\tau_L \approx 2.0104$}
\end{subfigure}\\[1ex]
\begin{subfigure}{1\linewidth}
\centering
\includegraphics[scale=0.48]{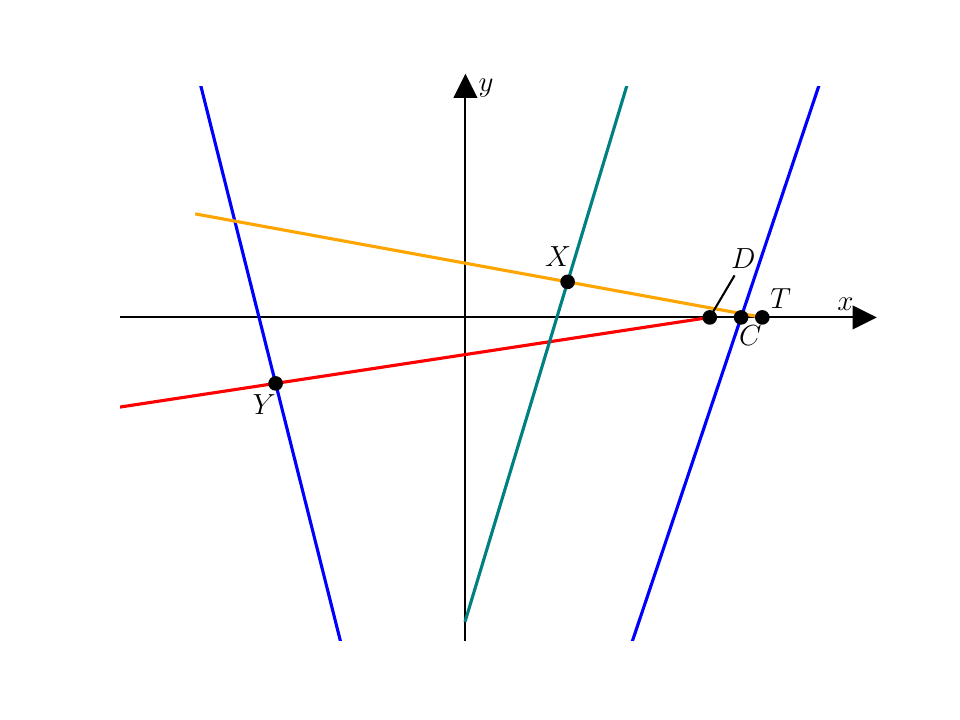}
\caption{$\tau_L = 2.2$}
\end{subfigure}
\caption{Phase portraits of \eqref{eq:BCNF2} with $\delta_L = -0.2$, $\delta_R=-0.2$, $\tau_R=-1.8$, and three different values of $\tau_L$
corresponding to the black triangles in Fig.~\ref{fig:reg2}-b.
Panel (b) uses $\tau_L$ such that $\phi^-(\xi) = 0$ to ten decimal places.
In (a) ${\rm cl}(W^u(X))$ is a chaotic attractor;
in (b) $W^u(X)$ and $W^s(Y)$ have corner intersections;
in (c) there is no attractor.}
\label{fig:typ_pp_2}
\end{figure}

Let 
\begin{align}
\label{eq:Phi2}
    \Phi^{(2)} = \left\{\xi \in \Phi\ \middle|\ \delta_L <0, \delta_R < 0\right\},
\end{align}
be the subset of $\Phi$ for which the BCNF is orientation-reversing.
As shown originally by Misiurewicz \cite{Mi80},
the closure of the unstable manifold of the fixed point $X$
can be a chaotic attractor.
This attractor contains the point $T$,
as shown in Fig.~\ref{fig:pp}-d, so under parameter variation is destroyed
when $T = C$.
This is a heteroclinic bifurcation beyond which
the unstable manifold of $X$ is unbounded.
From the formulas \eqref{eq:T} and \eqref{eq:C} we obtain
\begin{align}
\label{eq:CminusT}
C_1 - T_1 = \frac{\phi^-(\xi)}{(\lambda_L^u - 1)(1-\lambda_R^s)(\delta_R - \tau_R\lambda_L^u)},
\end{align}
where
\begin{align}
\label{eq:phi-}
\phi^-(\xi) = \delta_R - (\delta_R+\tau_R-(1+\lambda_R^u)\lambda_L^u)\lambda_L^u \,.
\end{align}
An example illustrating the destruction of the attractor
is shown in Fig.~\ref{fig:typ_pp_2}.
As the value of $\tau_L$ is increased the attractor
is destroyed when $T=C$ at $\tau_L \approx 2.0104$.


Analogous to the orientation-preserving case,
if $\delta_L = \delta_R = 0$ the expression $\phi^-(\xi)$ reduces to
$\phi^-(\xi) = \tau_L \phi_0(\tau_L,\tau_R)$.
Thus the heteroclinic bifurcation $\phi^-(\xi) = 0$
reduces to the familiar homoclinic bifurcation $\phi_0(\tau_L,\tau_R) = 0$
of the skew tent map family in the limit $(\delta_L,\delta_R) \to (0,0)$.

\begin{figure}[h]
\begin{tabular}{cc}
  \includegraphics[scale=0.5]{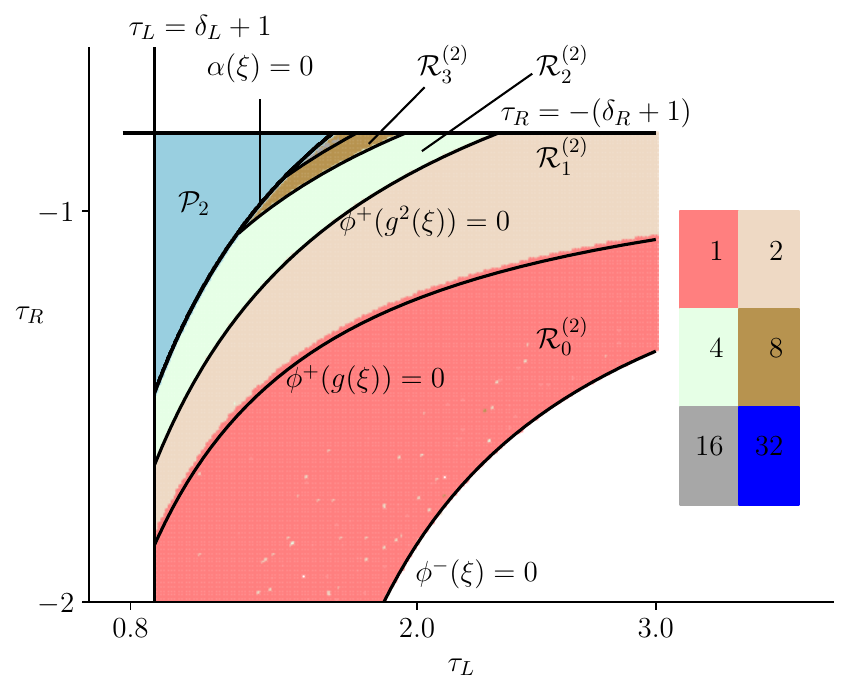} &   \includegraphics[scale=0.5]{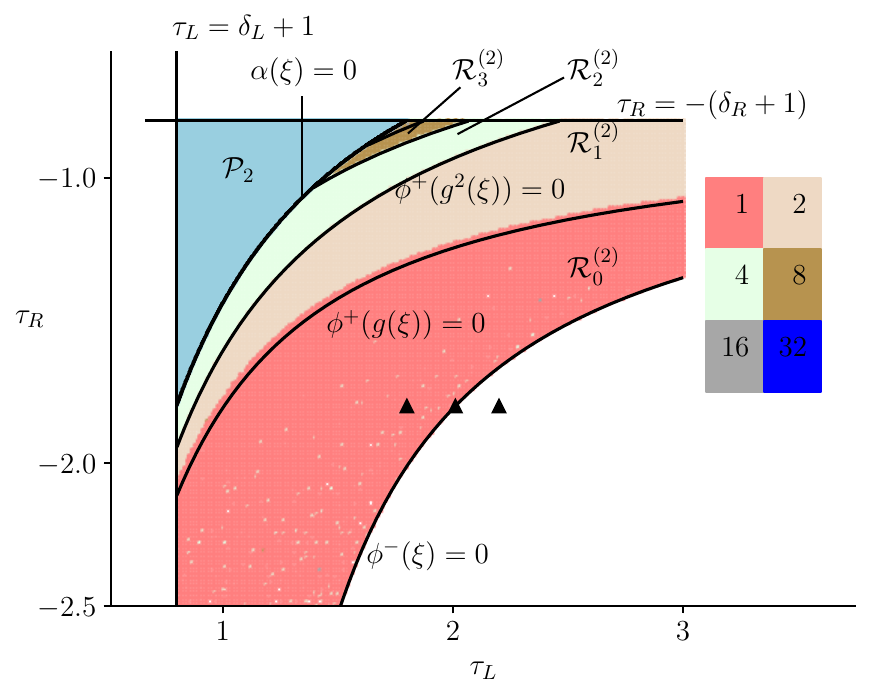}\\
(a) $\delta_L = -0.1, \delta_R=-0.2$. & (b) $\delta_L = -0.2, \delta_R=-0.2$.  \\[3pt]
\end{tabular}
\caption{
Two-dimensional slices of the
orientation-reversing parameter region $\Phi^{(2)}$
showing the bifurcation curves
overlaid upon numerical results
obtained using $\ee = 0.0001$ and $M = 10^6$.
The black triangles correspond to the parameter values
of Fig.~\ref{fig:typ_pp_2}.
}
\label{fig:reg2}
\end{figure}

We are now ready to subdivide $\Phi^{(2)}$ into regions
based upon the renormalisation operator $g$, \eqref{eq:g_xi}.
But $\xi \in \Phi^{(2)}$ implies $g(\xi) \in \Phi^{(1)}$,
so we again use the preimages of $\phi^+(\xi) = 0$ under $g$
to define the region boundaries.
Specifically we let
\begin{align}
\cR^{(2)}_0 &= \left\{\xi \in \Phi^{(2)} \,\middle|\, \phi^-(\xi) > 0, \phi^+(g(\xi)) \le 0, \alpha(\xi) < 0 \right\},
\label{eq:R_0^2} \\
\cR^{(2)}_n &= \left\{\xi \in \Phi^{(2)} \,\middle|\,
\phi^+ \left( g^n(\xi) \right) > 0,
\phi^+ \left( g^{n+1}(\xi) \right) \le 0,
\alpha(\xi) < 0 \right\}, \qquad \text{for all $n \ge 1$.}
\label{eq:R_n^2}
\end{align}
Fig.~\ref{fig:reg2} shows these regions in two typical
two-dimensional slices of parameter space.
Unlike in the orientation-preserving setting we need
to impose $\alpha(\xi) < 0$ in these regions
so that they don't overlap $\cP_2$ where a stable $LR$-cycle exists.

We conjecture that throughout $\cR^{(2)}_0$ the map has
a unique chaotic attractor with one connected component
equal to the closure of the unstable manifold of $X$, as in Fig.~\ref{fig:typ_pp_2}-a.
In earlier work \cite{GhMc23} we proved this to be true on a subset of $\cR^{(2)}_0$.
In the special case $\tau_L = -\tau_R$ and $\delta_L = \delta_R$ of the Lozi family of maps, the constraint $\phi^-(\xi) > 0$ reduces to
equation (3) of Misiurewicz \cite{Mi80}.

The following result is a simple consequence of Propositions
\ref{pr:0} and \ref{pr:gmaps} and \eqref{eq:R_n^2}.

\begin{proposition}
\label{pr:gmaps2}
If $\xi \in \cR_n^{(2)}$ with $n \ge 1$, then $g(\xi) \in \cR_{n-1}^{(1)}$.
\end{proposition}

\begin{figure}[h]
\begin{tabular}{cc}
  \includegraphics[scale=0.5]{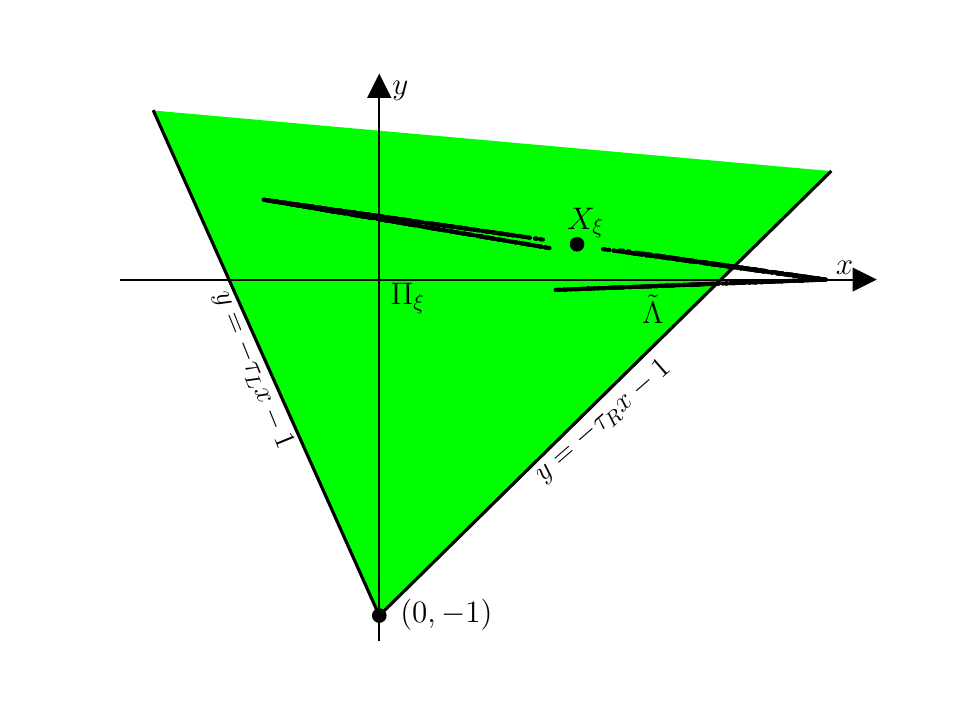} &   \includegraphics[scale=0.5]{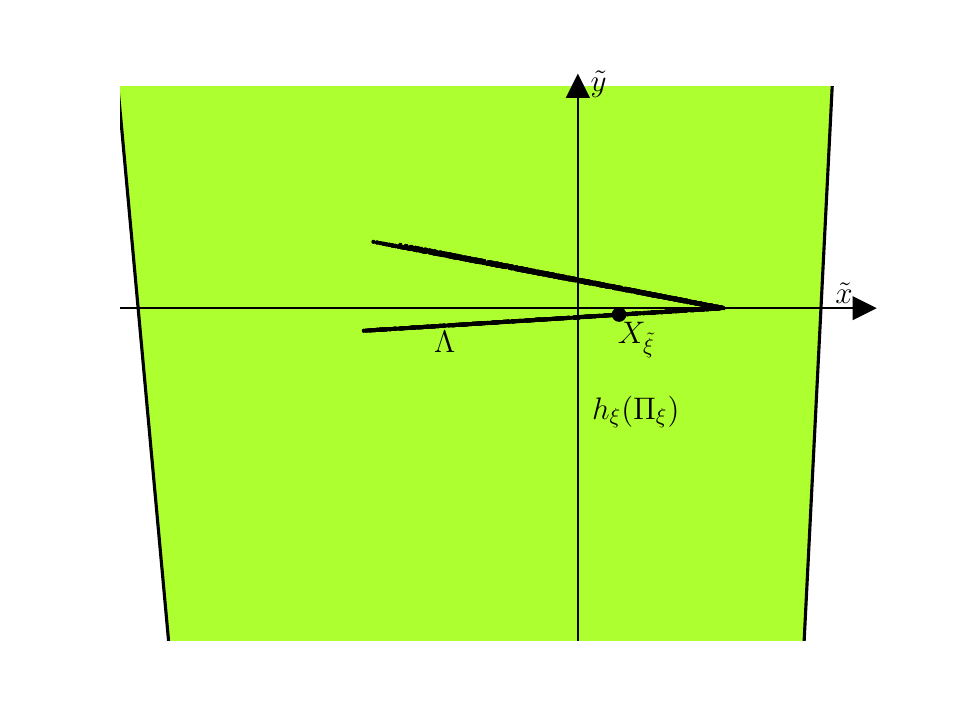}\\
(a) $\xi = \xi_{\rm ex}^{(2)} \in \mathcal{R}_1^{(2)}$ & (b) $\xi = g\big(\xi_{\rm ex}^{(2)}\big) \in \mathcal{R}_0^{(1)}$  \\[3pt]
\end{tabular}
\caption{
Panel (a) is a phase portrait of \eqref{eq:BCNF2}
at the parameter point $\xi_{\rm ex}^{(2)}$ \eqref{eq:xi2ex}
where the attractor has two connected components
in the orientation-reversing case.
Panel (b) uses instead $g\big(\xi_{\rm ex}^{(2)}\big)$.
}
\label{fig:att_2}
\end{figure}

This suggests that for any $\xi \in \cR_n^{(2)}$
the BCNF has an attractor with $2^n$ connected components.
For example 
\begin{align}
    \xi_{\rm ex}^{(2)} = \left(2.5, -0.1, -1.1, -0.2 \right)
    \label{eq:xi2ex}
\end{align}
belongs to $\cR^{(2)}_1$ and indeed the attractor shown
in Fig.~\ref{fig:att_2}-a appears to have two connected components.
One component belongs to $\Pi_\xi$
so Proposition \ref{pr:C} applies.
Hence this component is an affine transformation
of the attractor of \eqref{eq:BCNF2} at the parameter point
$g \big( \xi_{\rm ex}^{(2)} \big)$ that belongs to $\cR^{(1)}_0$, and we know its attractor has one component by Theorem \ref{th:gen_R0}.

Fig.~\ref{fig:reg2} shows that our conjecture
is supported by the output of the greatest common divisor algorithm.
The boundaries of the $\cR^{(2)}_n$ closely approximate
the places with the value of the greatest common divisor changes.
This value changes slightly above $\phi^+(g(\xi)) = 0$
because here the two components of the attractor are very close
and $\ee = 0.0001$ is insufficient to detect this difference.
Also $\cR^{(2)}_0$ has pixels erroneously corresponding to more than one component because here the attractor is relatively large
and $M = 10^6$ iterations are insufficient
for the algorithm to consistently obtain a greatest common divisor of $1$.

\section{The non-invertible case $\delta_L > 0$, $\delta_R < 0$.}
\label{sec:nonInvertNice}

\begin{figure}
\begin{subfigure}{.5\linewidth}
\centering
\includegraphics[scale=0.48]{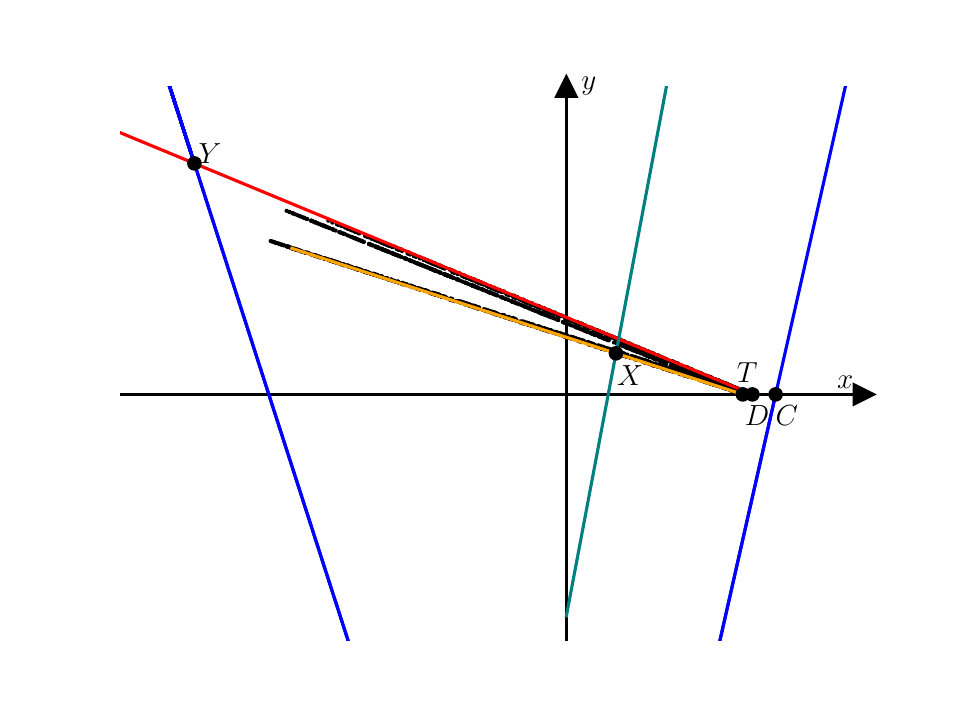}
\caption{$\tau_L = 1.7$}
\end{subfigure}%
\begin{subfigure}{.5\linewidth}
\centering
\includegraphics[scale=0.48]{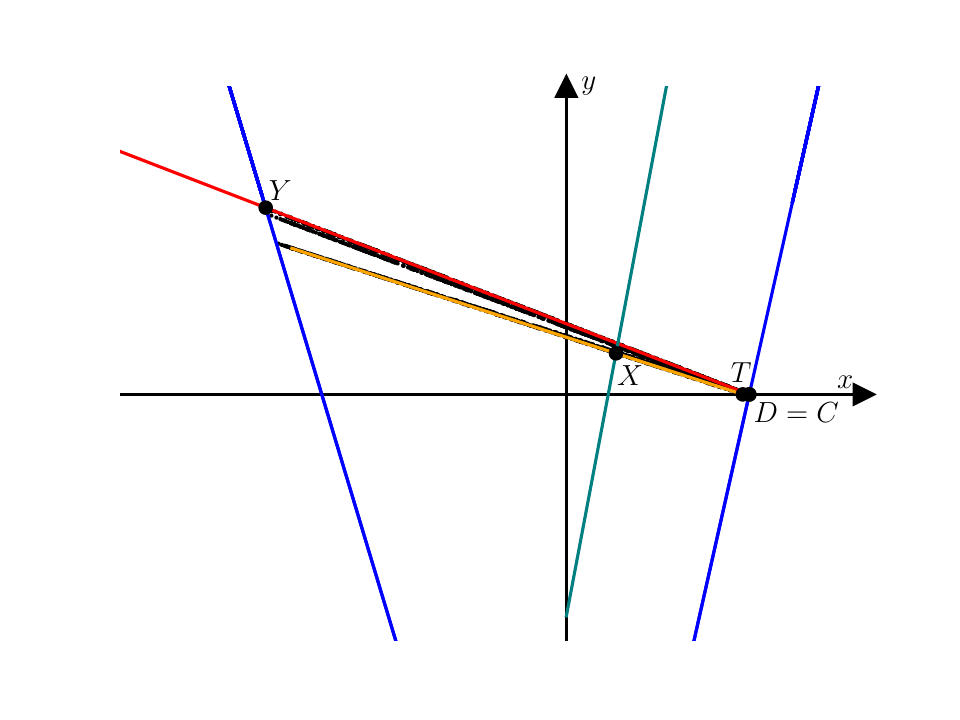}
\caption{$\tau_L \approx 1.7948$}
\end{subfigure}\\[1ex]
\begin{subfigure}{1\linewidth}
\centering
\includegraphics[scale=0.48]{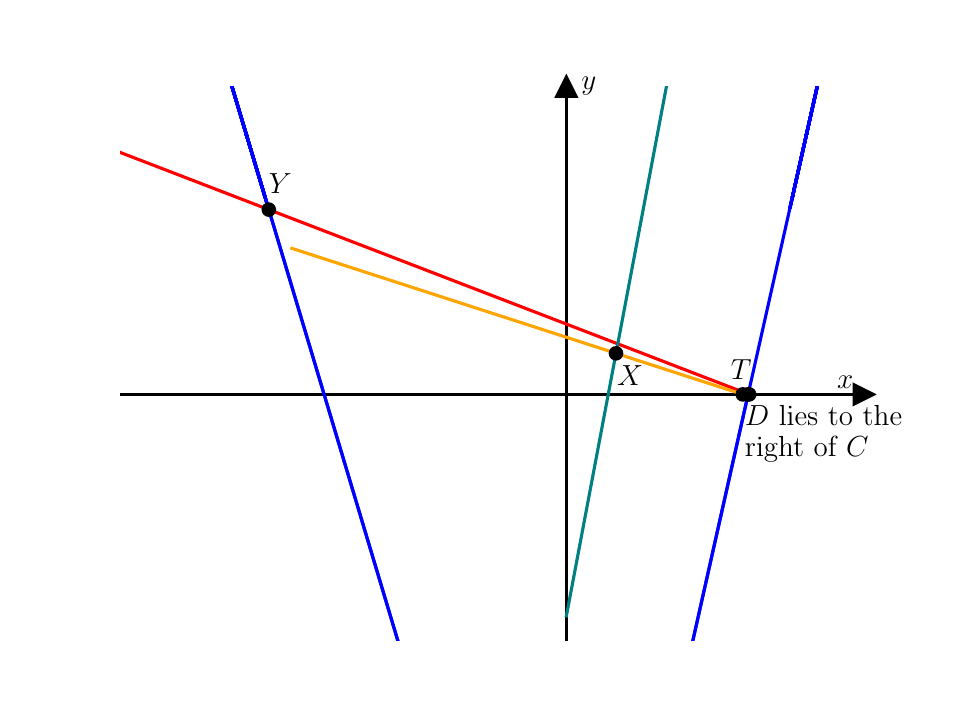}
\caption{$\tau_L = 1.8$}
\end{subfigure}
\caption{Phase portraits of \eqref{eq:BCNF2} with $\delta_L = 0.3$, $\delta_R=-0.4$, $\tau_R=-2.4$ and three different values of $\tau_L$ corresponding to the blue triangles in Fig.~\ref{fig:reg3}-b.}
\label{fig:typ_pp_3B}
\end{figure}

\begin{figure}
\begin{subfigure}{.5\linewidth}
\centering
\includegraphics[scale=0.48]{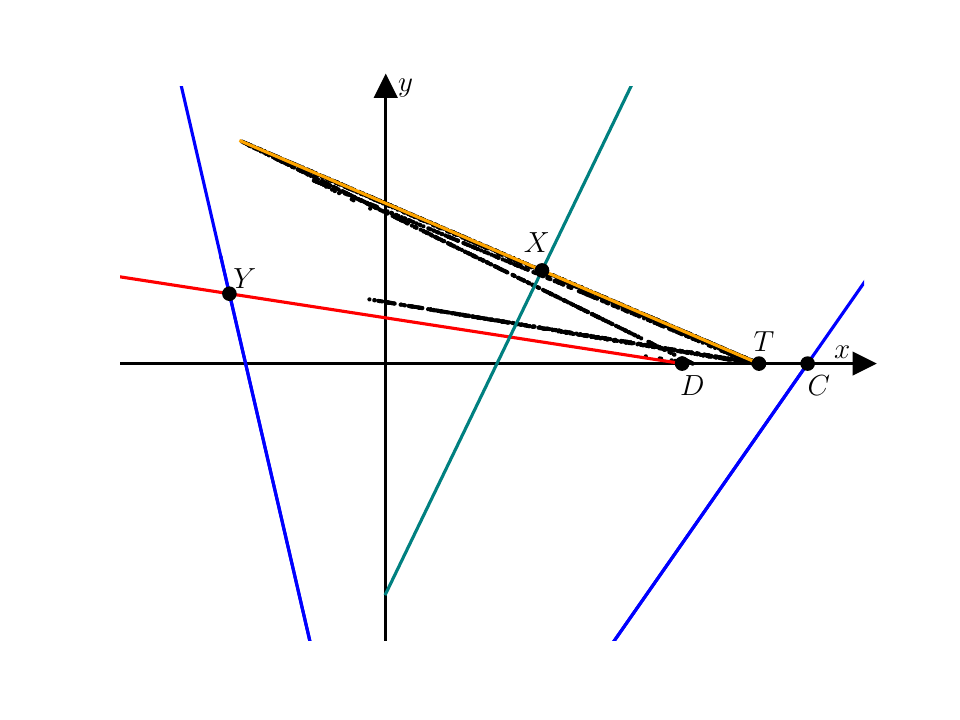}
\caption{$\tau_R = -1.1$}
\end{subfigure}%
\begin{subfigure}{.5\linewidth}
\centering
\includegraphics[scale=0.48]{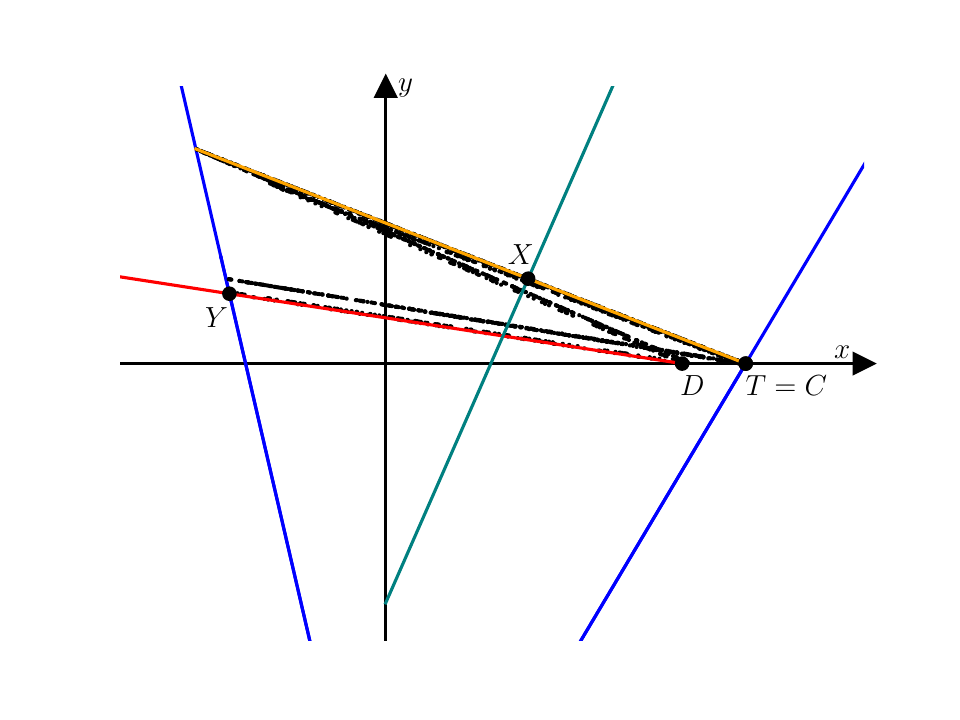}
\caption{$\tau_R \approx -1.2654$}
\end{subfigure}\\[1ex]
\begin{subfigure}{1\linewidth}
\centering
\includegraphics[scale=0.48]{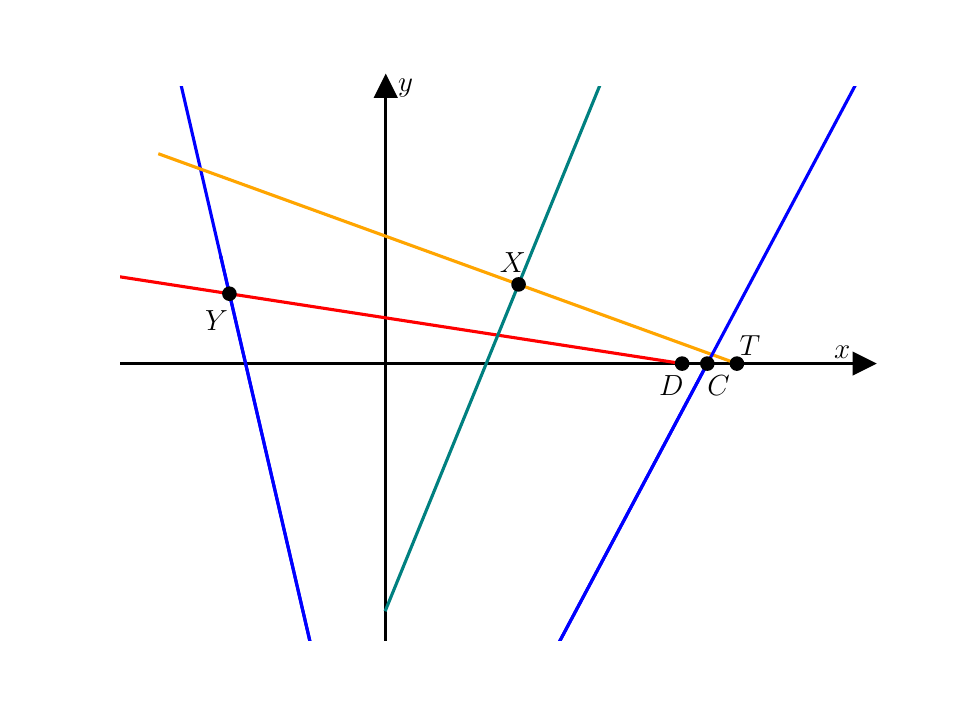}
\caption{$\tau_R = -1.4$}
\end{subfigure}
\caption{Phase portraits of \eqref{eq:BCNF2} with $\delta_L = 0.3$, $\delta_R=-0.4$, $\tau_L=3$, and three different values of $\tau_R$ corresponding to the black triangles in Fig.~\ref{fig:reg3}-b.}
\label{fig:typ_pp_3A}
\end{figure}

\begin{figure}[h]
\begin{tabular}{cc}
  \includegraphics[scale=0.5]{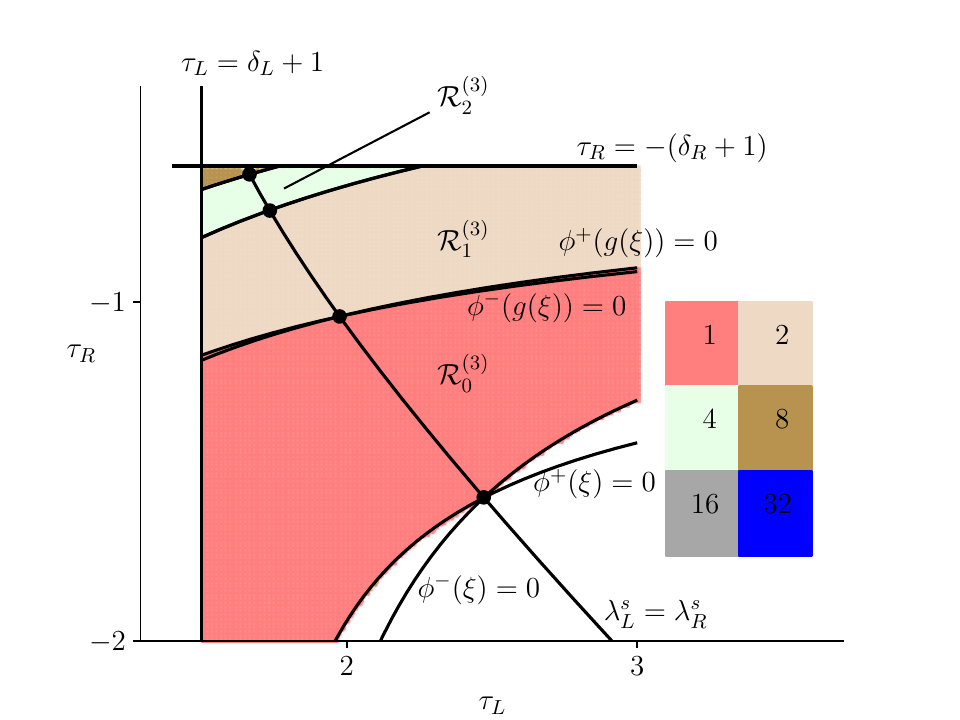} &   \includegraphics[scale=0.5]{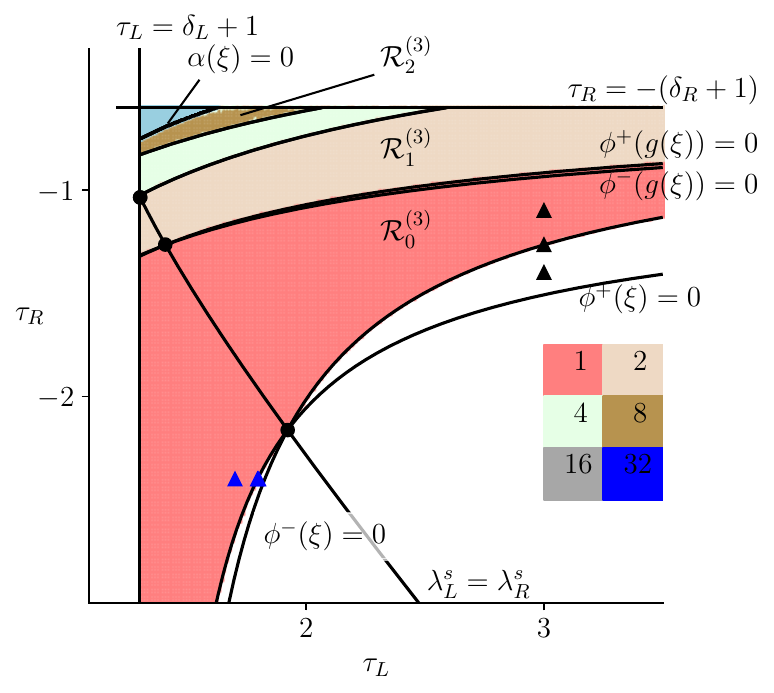}\\
(a) $\delta_L = 0.5, \delta_R=-0.4$. & (b) $\delta_L = 0.3, \delta_R=-0.4$.  \\[3pt]
\end{tabular}
\caption{
Two-dimensional slices of the
non-invertible parameter region $\Phi^{(3)}$
showing the bifurcation curves
overlaid upon numerical results
obtained using $\ee = 0.001$ and $M = 10^6$.
The blue [black] triangles correspond to the parameter values
of Fig.~\ref{fig:typ_pp_3B} [Fig.~\ref{fig:typ_pp_3A}].
}
\label{fig:reg3}
\end{figure}

Here we study the parameter region
\begin{align}
\label{eq:Phi3}
    \Phi^{(3)} = \left\{\xi \in \Phi\ \middle|\ \delta_L >0, \delta_R < 0\right\},
\end{align}
where \eqref{eq:BCNF2} is non-invertible.
In this region an attractor of \eqref{eq:BCNF2}
can be destroyed by crossing the homoclinic bifurcation
$\phi^+(\xi) = 0$ or the heteroclinic bifurcation $\phi^-(\xi) = 0$.
This is because near these boundaries the attractor
contains the point $T$ and is close to the point $D$
so is destroyed when one of these points collides with $C$.
For example in Fig.~\ref{fig:typ_pp_3B} $T$ lies to the left of $D$,
so the attractor is destroyed when $D = C$, i.e.~when $\phi^+(\xi) = 0$.
In contrast in Fig.~\ref{fig:typ_pp_3A} $T$ lies to the right of $D$,
so the attractor is destroyed when $T = C$, i.e.~when $\phi^-(\xi) = 0$.
For this reason we define
\begin{align}
\label{eq:phi_min}
\phi_{\rm min}(\xi) = {\rm min}[\phi^+(\xi), \phi^-(\xi)],
\end{align}
and
\begin{align}
\label{eq:Rn3}
    \cR^{(3)}_n = \left\{ \xi \in \Phi^{(3)} \,\middle|\, 
    \phi_{\rm min} \left( g^n(\xi) \right) > 0,\,
    \phi_{\rm min} \left( g^{n+1}(\xi) \right) \le 0,\,
    \alpha(\xi) < 0 \right\},
\end{align}
for all $n \ge 0$.
Two-dimensional slices of these regions are shown in Fig.~\ref{fig:reg3}.
We conjecture that throughout the first region $\cR_0^{(3)}$ the BCNF has a unique chaotic attractor
with one connected component, and this is supported by the numerical results shown in Fig.~\ref{fig:reg3}.
To explain the dynamics in the remaining regions we use the following analogy
to Propositions \ref{pr:gmaps1} and \ref{pr:gmaps2}.
Here, however, the result is not trivial, so we provide a proof.


\begin{proposition}
\label{pr:gmaps3}
If $\xi \in \cR^{(3)}_n$ with $n \ge 1$, then $g(\xi) \in \cR^{(3)}_{n-1}$.
\end{proposition}

\begin{proof}
Choose any $\xi \in \cR^{(3)}_n$ with $n \ge 1$
and write $g(\xi) = \big( \tilde{\tau}_L, \tilde{\delta}_L, \tilde{\tau}_R, \tilde{\delta}_R \big)$ as in the proof of Proposition \ref{pr:gmaps}.
By Proposition \ref{pr:gmaps} we have $g(\xi) \in \Phi$.
Further, $g(\xi) \in \Phi^{(3)}$ because
$\tilde{\delta}_L = \delta_R^2 > 0$ and $\tilde{\delta}_R = \delta_L \delta_R < 0$.
Also $\phi_{\rm min} \big( g^{n-1}(g(\xi)) \big) > 0$
and $\phi_{\rm min} \big( g^n(g(\xi)) \big) \le 0$.
Finally $\alpha(g(\xi)) = \tilde{\tau}_L \tilde{\tau}_R + (\tilde{\delta}_L - 1)(\tilde{\delta}_R - 1)$ where
$\tilde{\tau}_L = \tau_R^2 - 2 \delta_R$,
$\tilde{\delta}_L = \delta_R^2$,
$\tilde{\tau}_R = \tau_L \tau_R - \delta_L - \delta_R$, and
$\tilde{\delta}_R = \delta_L \delta_R$.
By substituting $\tau_L > \delta_L + 1$ and $\tau_R < -(\delta_R + 1)$
(true because $\xi \in \Phi$) we obtain (after simplification)
$$
\alpha(g(\xi)) < -2 \left( 1 + \delta_R + \delta_R^2 \right)(\delta_L + \delta_R).
$$
So if $\delta_L + \delta_R > 0$ then $\alpha(g(\xi)) < 0$
(because $1 + \delta_R + \delta_R^2 \ge \frac{3}{4} > 0$).
If instead $\delta_L + \delta_R \le 0$ then
$\tilde{\delta}_L + \tilde{\delta}_R = \delta_R (\delta_L + \delta_R) \ge 0$ because $\delta_R < 0$, so again $\alpha(g(\xi)) < 0$
by Proposition \ref{pr:0} applied to the parameter point $g(\xi)$.
\end{proof}

\begin{figure}[h]
\begin{tabular}{cc}
  \includegraphics[scale=0.5]{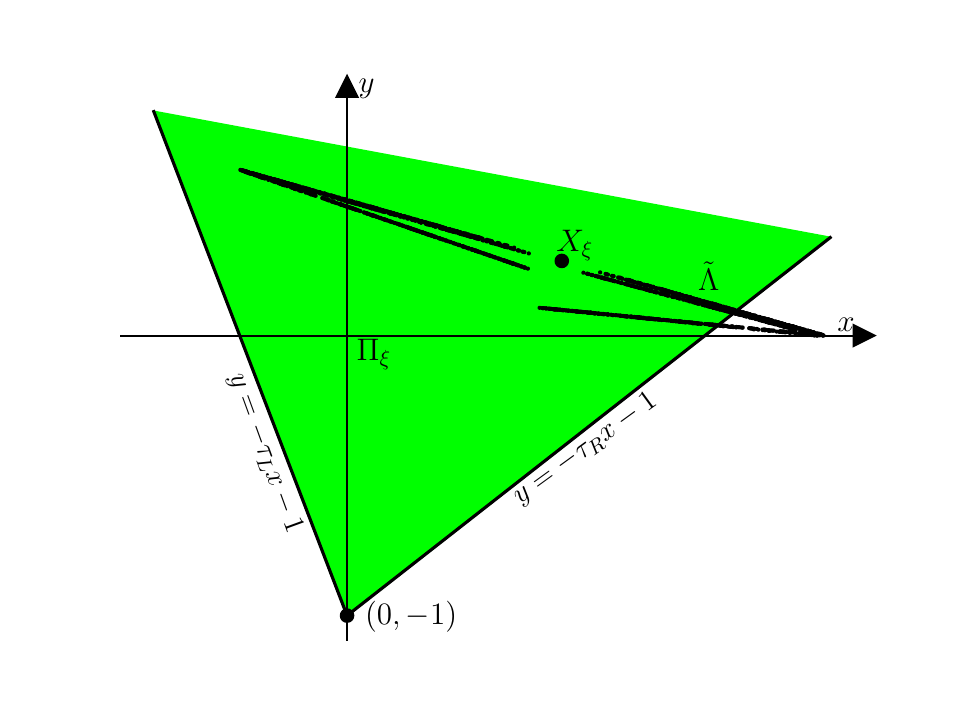} &   \includegraphics[scale=0.5]{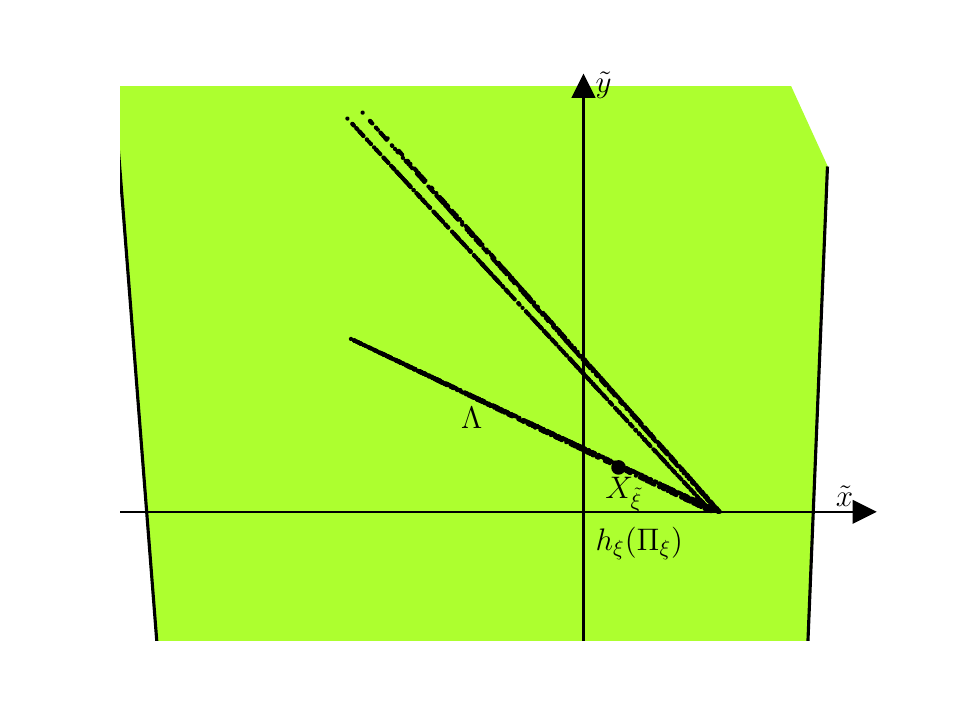}\\
(a) $\xi = \xi_{\rm ex}^{(3)} \in \mathcal{R}_1^{(3)}$ & (b) $\xi = g\big(\xi_{\rm ex}^{(3)}\big) \in \mathcal{R}_0^{(3)}$  \\[3pt]
\end{tabular}
\caption{
Panel (a) is a phase portrait of \eqref{eq:BCNF2}
at the parameter point $\xi_{\rm ex}^{(3)}$ \eqref{eq:xi3ex}
where the attractor has two connected components
in the $\delta_L > 0$ non-invertible case.
Panel (b) uses instead $g\big(\xi_{\rm ex}^{(3)}\big)$.
}
\label{fig:att_3}
\end{figure}

Proposition \ref{pr:gmaps3} suggests that throughout each $\cR^{(3)}_n$ with $n \ge 1$
the BCNF has an attractor with exactly $2^n$ connected components,
and Fig.~\ref{fig:reg3} supports this conjecture.
Fig.~\ref{fig:att_3}-a provides an example using the parameter point
\begin{align}
    \xi_{\rm ex}^{(3)} = \left(3, 0.3, -.9, -0.4\right),
    \label{eq:xi3ex}
\end{align}
which belongs to $\cR_1^{(3)}$.
The attractor has two pieces, one of which belongs to $\Pi_\xi$,
so as above this piece is an affine transformation of the attractor of $g \big( \xi^{(3)}_{\rm ex} \big) \in \cR^{(3)}_0$
shown in Fig.~\ref{fig:att_3}-b.

\section{The non-invertible case $\delta_L < 0$, $\delta_R > 0$}
\label{sec:nonInvertNotNice}

It remains for us to consider
\begin{align}
\label{eq:Phi4}
    \Phi^{(4)} = \left\{\xi \in \Phi\ \middle|\ \delta_L <0, \delta_R > 0\right\},
\end{align}
where \eqref{eq:BCNF2} is non-invertible.
In this region the attractor is usually destroyed before the boundaries $\phi^+(\xi) = 0$ and $\phi^-(\xi) = 0$
in a heteroclinic bifurcation
that cannot be characterised by an explicit condition on the parameter values.
This occurs when the attractor contains points on $y=0$ that lie to the right of $D$ and $T$
and is destroyed when its right-most point collides with $C$.

\begin{figure}
\begin{subfigure}{.5\linewidth}
\centering
\includegraphics[scale=0.48]{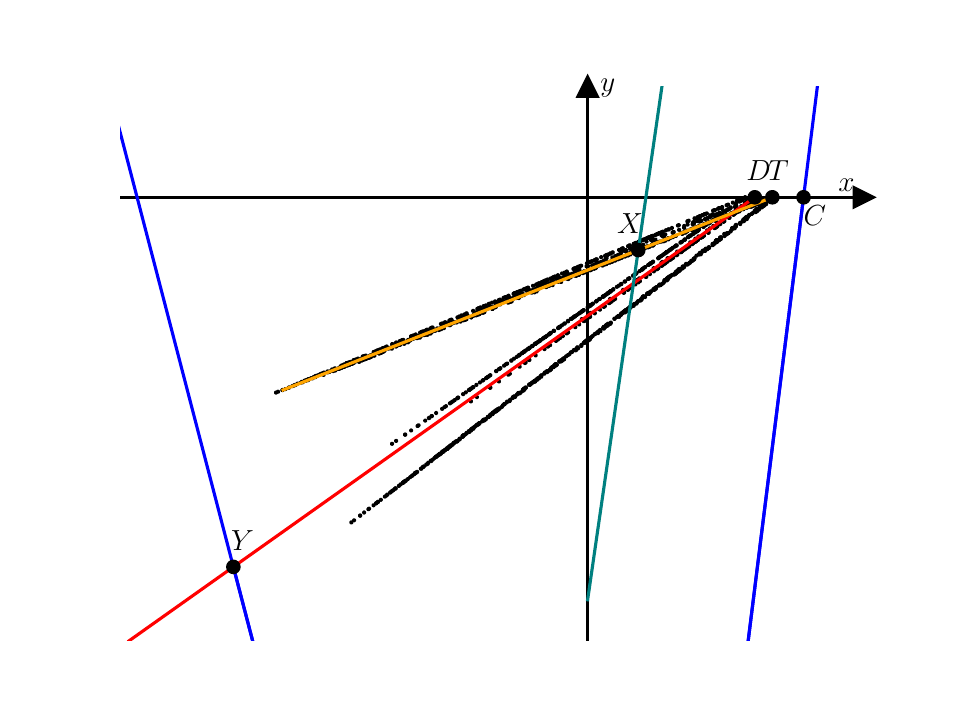}
\caption{$\tau_L = 1.2$}
\end{subfigure}%
\begin{subfigure}{.5\linewidth}
\centering
\includegraphics[scale=0.48]{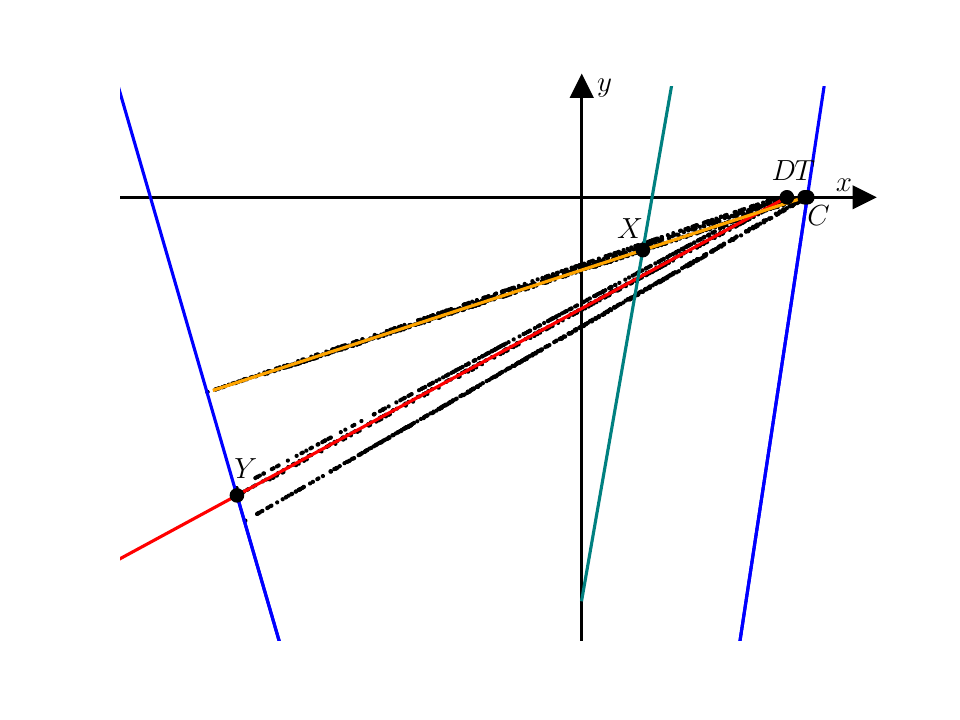}
\caption{$\tau_L \approx 1.3439$}
\end{subfigure}\\[1ex]
\begin{subfigure}{1\linewidth}
\centering
\includegraphics[scale=0.48]{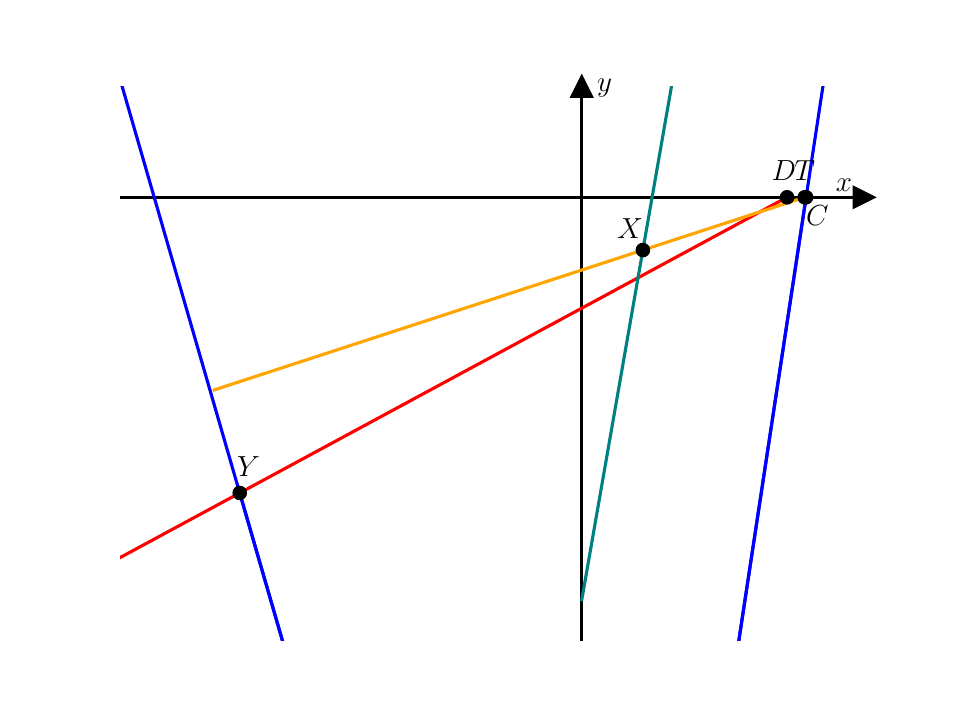}
\caption{$\tau_L = 1.35$}
\end{subfigure}
\caption{Phase portraits of \eqref{eq:BCNF2} with $\delta_L = -0.4$, $\delta_R=0.4$, $\tau_R=-2.8$, and three different values of $\tau_L$ corresponding to the blue triangles in Fig.~\ref{fig:reg4}-a.
The parameter value used in panel (b) is approximately
where the attractor is destroyed.
In all three panels $D$ lies to the left of $T$
which lies to the left of $C$.}
\label{fig:typ_pp_4B}
\end{figure}

\begin{figure}
\begin{subfigure}{.5\linewidth}
\centering
\includegraphics[scale=0.48]{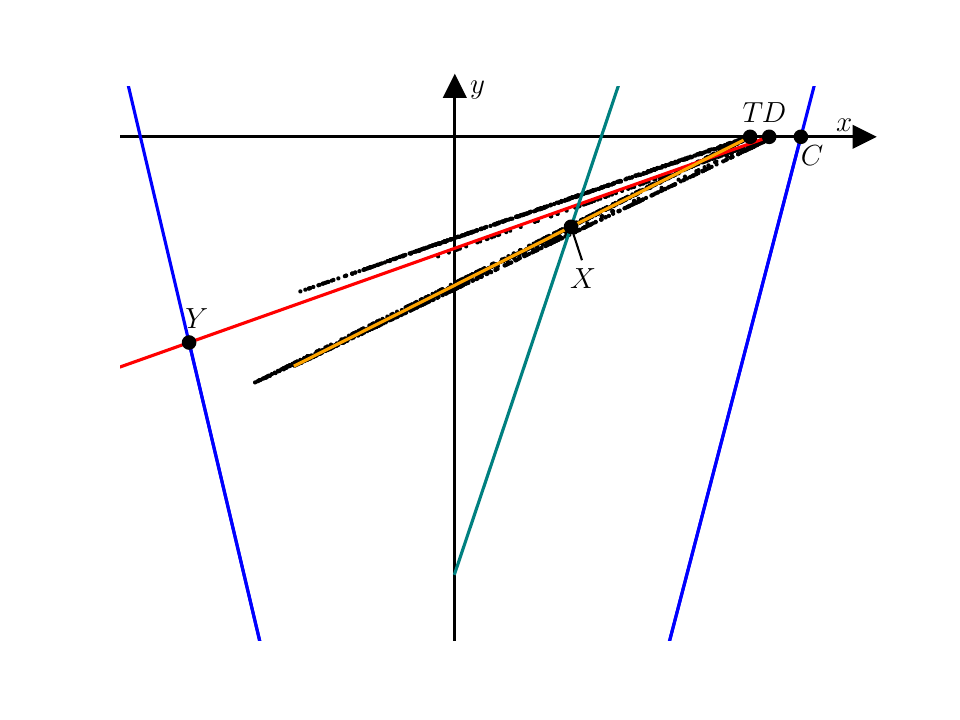}
\caption{$\tau_L = 2$}
\end{subfigure}%
\begin{subfigure}{.5\linewidth}
\centering
\includegraphics[scale=0.48]{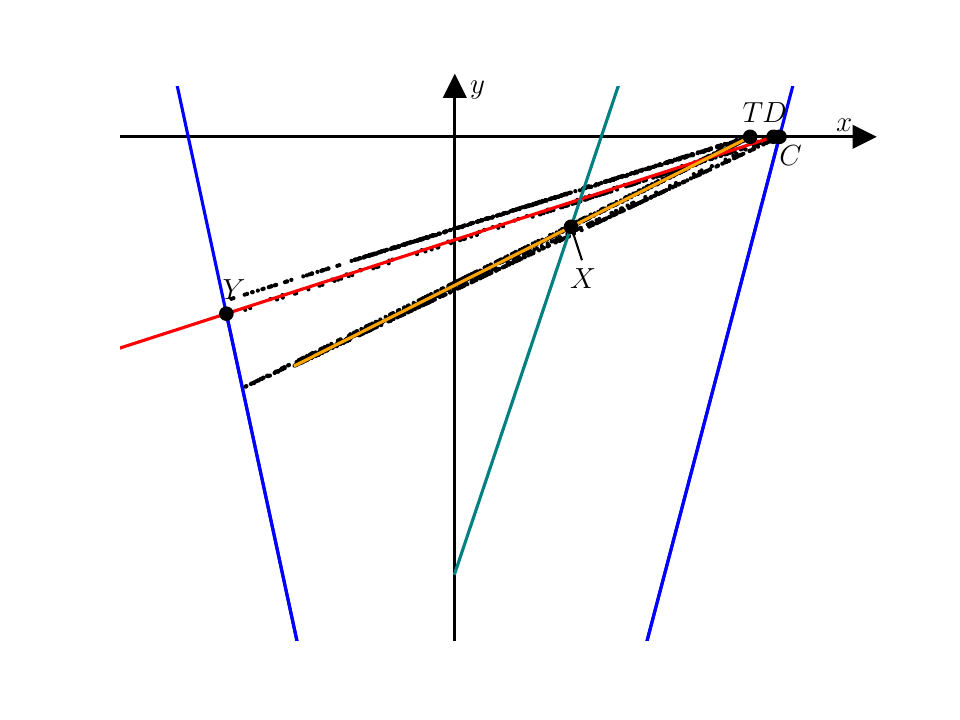}
\caption{$\tau_L \approx 2.2285$}
\end{subfigure}\\[1ex]
\begin{subfigure}{1\linewidth}
\centering
\includegraphics[scale=0.48]{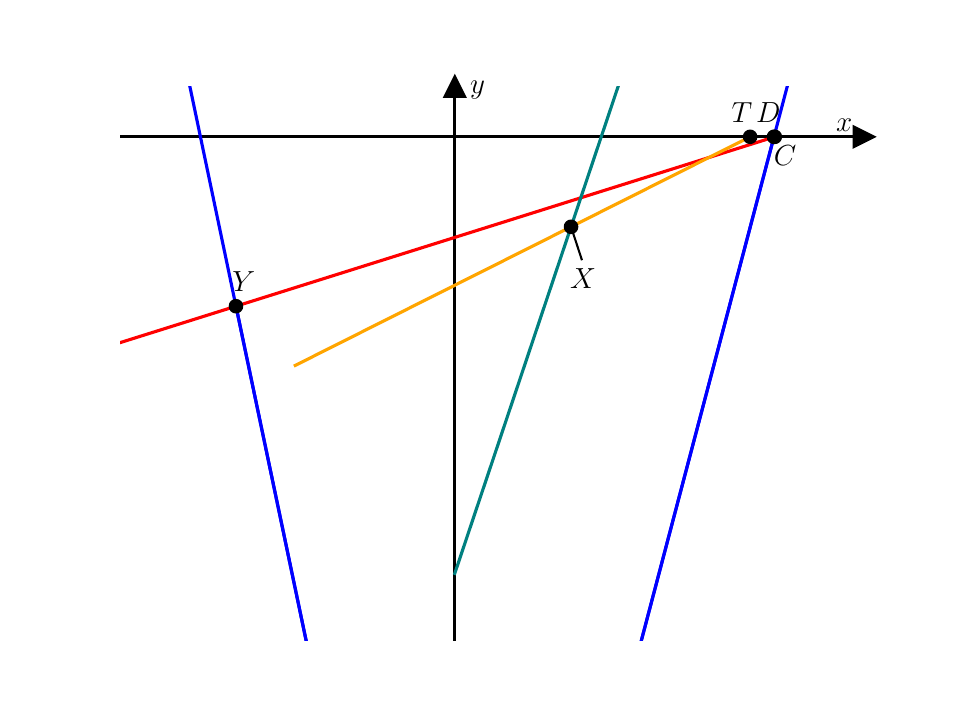}
\caption{$\tau_L = 2.3$}
\end{subfigure}
\caption{Phase portraits of \eqref{eq:BCNF2} with $\delta_L = -0.4$, $\delta_R=0.4$, $\tau_R=-1.8$, and three different values of $\tau_L$ corresponding to the black triangles in Fig.~\ref{fig:reg4}-a.
The parameter value used in panel (b) is approximately
where the attractor is destroyed.
In all three panels $T$ lies to the left of $D$
which lies to the left of $C$.}
\label{fig:typ_pp_4A}
\end{figure}

Fig.~\ref{fig:typ_pp_4B} shows an example.
In panel (a) the attractor is the closure of the unstable manifold of $X$.
As we grow the unstable manifold outwards from $X$ it develops points on $y=0$
that lie further and further to the right, but not beyond $C$.
This occurs as a consequence of the geometric configuration afforded by $f_R$ being orientation-preserving
and $f_L$ being orientation-reversing.
As the value of $\tau_L$ is increased the attractor is destroyed
when its right-most point collides with $C$, panel (b).
With a slightly value of $\tau_L$, panel (c),
typical forward orbits diverge even though $D$ and $T$ still lie to the left of $C$.
Fig.~\ref{fig:typ_pp_4A} provides a second example.
Here the parameter values are such that $D$ and $T$ are switched around
but the attractor is destroyed in the same way.


\begin{figure}
\begin{tabular}{cc}
  \includegraphics[scale=0.5]{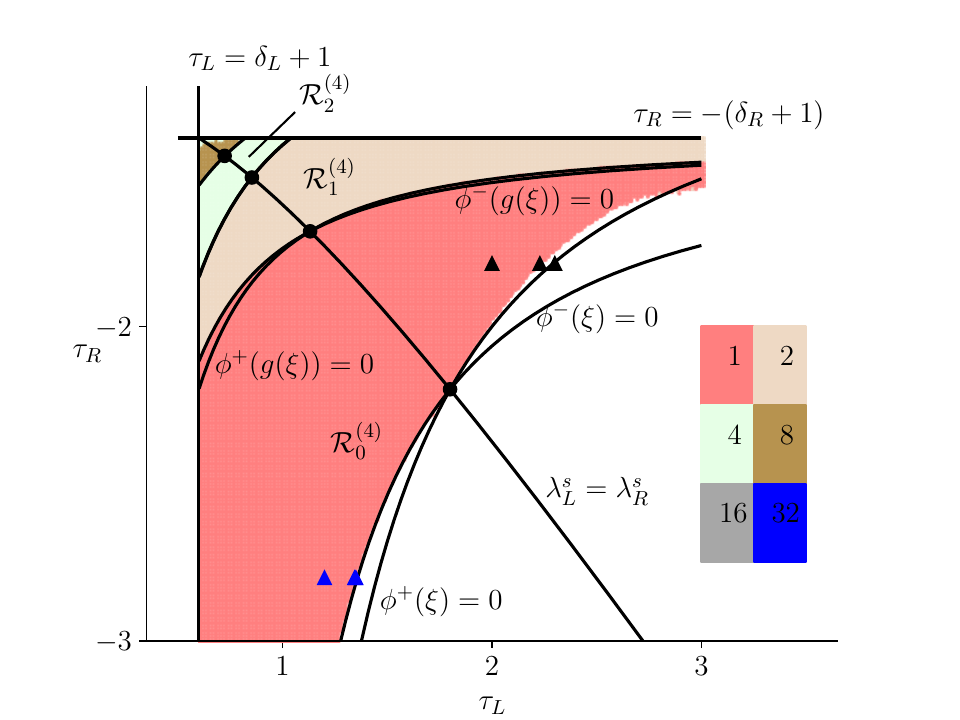} &   \includegraphics[scale=0.5]{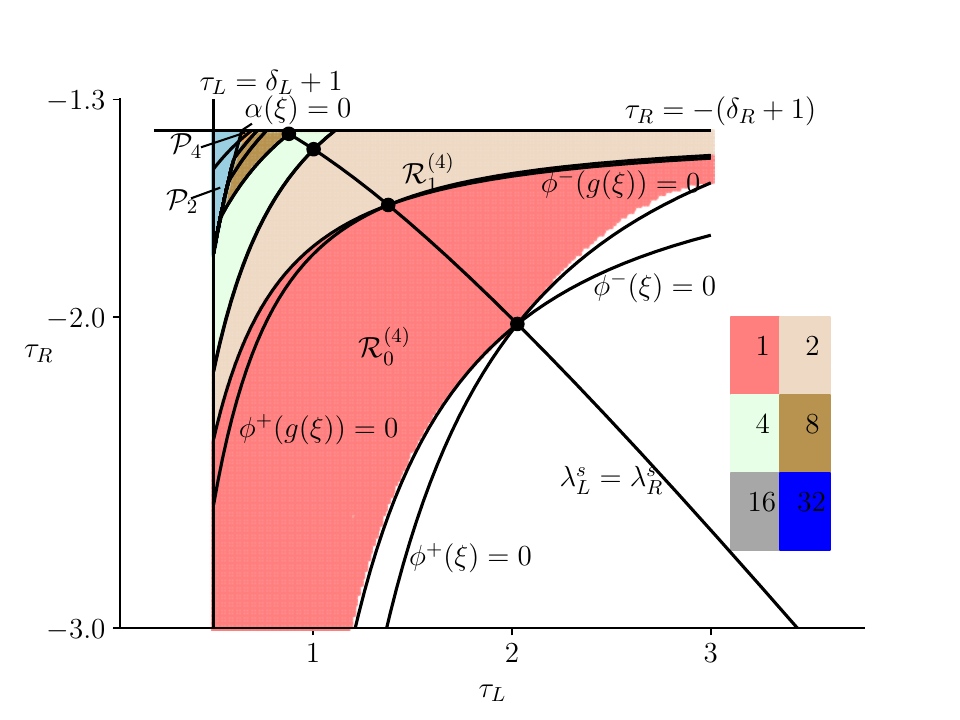}\\
(a) $\delta_L = -0.4, \delta_R=0.4$. & (b) $\delta_L = -0.5, \delta_R=0.4$.  \\[3pt]
\end{tabular}
\caption{
Two-dimensional slices of the
non-invertible parameter region $\Phi^{(4)}$
showing the bifurcation curves
overlaid upon numerical results
obtained using $\ee = 0.001$ and $M = 10^6$.
The blue [black] triangles correspond to the parameter values
of Fig.~\ref{fig:typ_pp_4B} [Fig.~\ref{fig:typ_pp_4A}].
}
\label{fig:reg4}
\end{figure}

Fig.~\ref{fig:reg4} shows the results of numerical simulations applied to two slices of $\Phi^{(4)}$.
The white areas just to the left of $\phi^+(\xi) = 0$ and $\phi^-(\xi) = 0$
correspond to the phenomenon that we have just described.
These areas are bounded on the left by a curve of heteroclinic bifurcations.
Since we cannot identify this curve with hand calculations,
it is natural to attempt to compute the curve numerically with numerical continuation methods.
However, we do not show the result of such a computation
because we suspect this curve highly irregular, e.g.~non-differentiable at infinitely many points,
as explained by Osinga \cite{Os06}.

In each plot in Fig.~\ref{fig:reg4} the attractor does in fact persist up to where
the boundaries $\phi^+(\xi) = 0$ and $\phi^-(\xi) = 0$ meet.
This is because on the curve $\lambda_L^s = \lambda_R^s$
the right-most point of the attractor is $D$ and $T$, which are equal.
Also near the top of Fig.~\ref{fig:reg4}-a the attractor persists slightly beyond $\phi^+(\xi) = 0$
because here the attractor fails to approach $D$ so is not destroyed when $D=C$ at $\phi^+(\xi) = 0$.

\begin{figure}[]
\vskip 6pt
\centering
\includegraphics[width=0.8\linewidth]{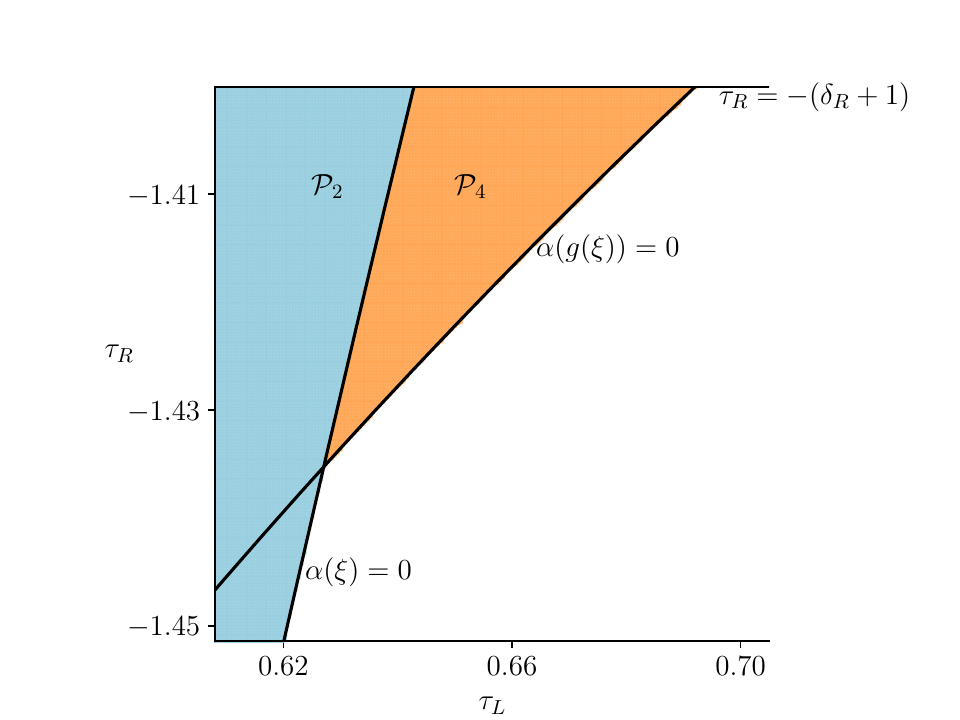}
\caption{A magnified version of Fig.~\ref{fig:reg4}-b showing the regions having a stable $LR$-cycle (period-two solution) and a stable $LRRR$-cycle (period-four solution).
}\label{fig:mag}
\end{figure}

Despite the extra complexities in $\Phi^{(4)}$ it still appears that renormalisation
is helpful for explaining the bifurcation structure.
Let
\begin{align}
\label{eq:R_0^4}
    \cR^{(4)}_0 = \left\{\xi \in \Phi^{(4)} \middle|\ \phi_{\rm min}(\xi)>0,\, \phi_{\rm min}(g(\xi)) \le 0,\,
    \alpha(\xi) < 0 \right\}.
\end{align}
Based on Fig.~\ref{fig:reg4} we conjecture that for any $\xi \in \cR^{(4)}_0$,
if \eqref{eq:BCNF2} has an attractor then this attractor is chaotic with one connected component.
Now let
\begin{align}
\label{Rn4}
\cR^{(4)}_n = \left\{\xi \in \Phi^{(4)} \middle|\ \phi_{\rm min}(g^n(\xi)) >0,\, \phi_{\rm min}(g^{n+1}(\xi)) \le 0,\,
\alpha(\xi) < 0, \alpha(g(\xi)) < 0 \right\}.
\end{align}
Unlike in previous sections here we have included the extra constraint $\alpha(g(\xi)) < 0$
so that the $\cR^{(4)}_n$ do not include the region $\mathcal{P}_4$,
defined to be where \eqref{eq:BCNF2} has a stable period-four solution with symbolic itinerary $LRRR$.
This region is visible in Fig.~\ref{fig:reg4}-b and shown more clearly in the magnification, Fig.~\ref{fig:mag}. Note that here we do not show the result of the numerics because the component counting algorithm does not work efficiently close to the curve $\alpha(g(\xi))=0$. The following result is a trivial consequence of our definitions.

\begin{proposition}
\label{pr:gmaps4}
If $\xi \in \cR^{(4)}_n$ with $n \ge 1$, then $g(\xi) \in \cR^{(3)}_{n-1}$.
\end{proposition}

\begin{figure}[h]
\begin{tabular}{cc}
  \includegraphics[scale=0.5]{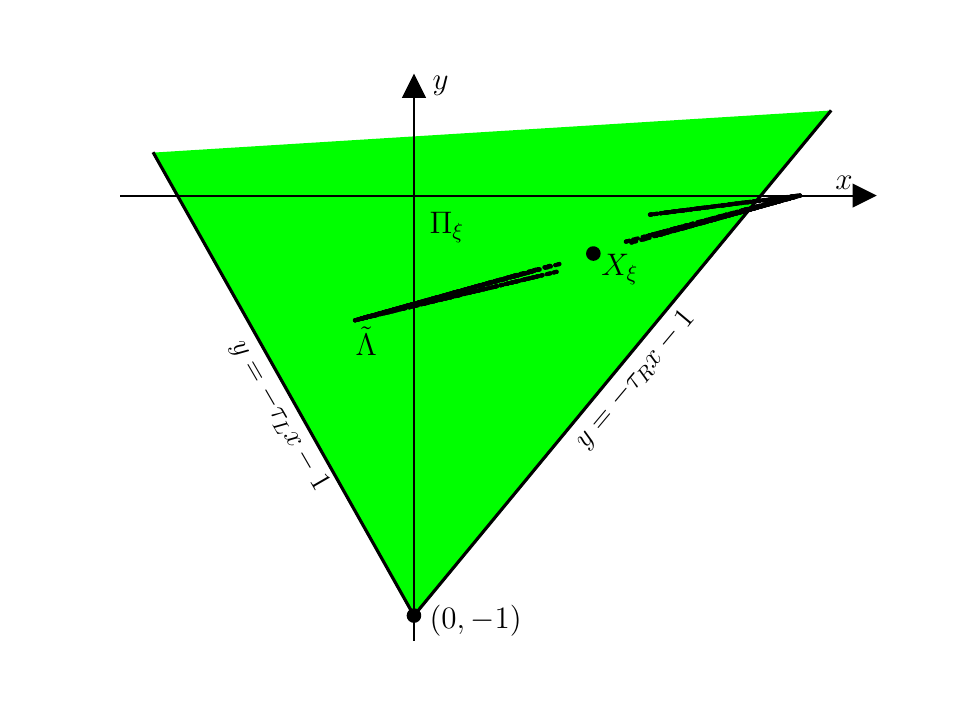} &   \includegraphics[scale=0.5]{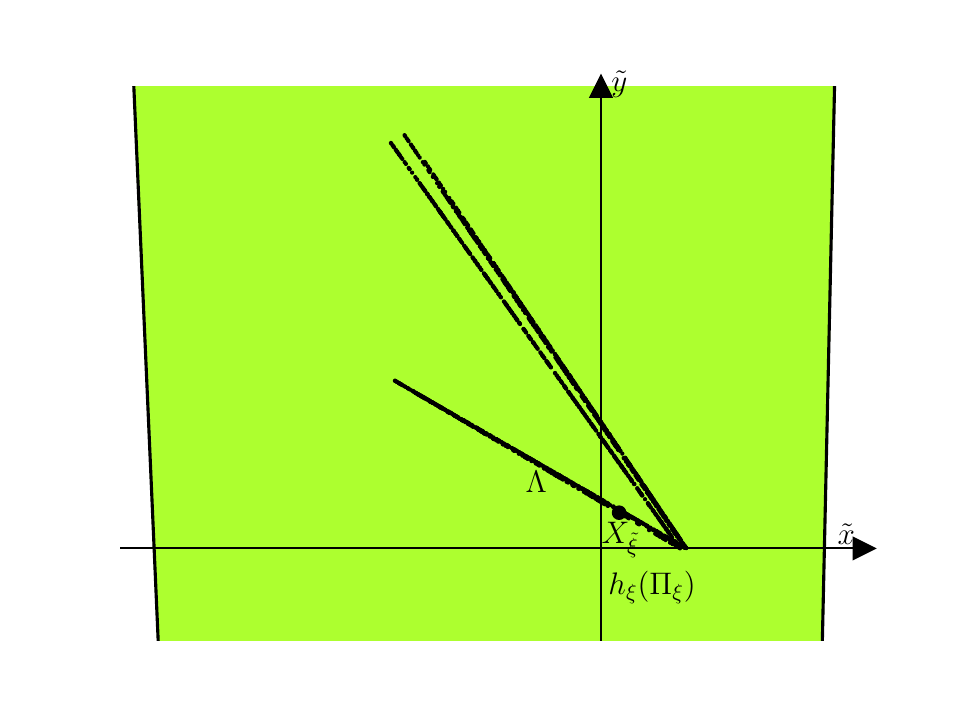}\\
(a) $\xi = \xi_{\rm ex}^{(4)} \in \mathcal{R}_1^{(4)}$ & (b) $\xi = g\big(\xi_{\rm ex}^{(4)}\big) \in \mathcal{R}_0^{(3)}$  \\[3pt]
\end{tabular}
\caption{
Panel (a) is a phase portrait of \eqref{eq:BCNF2}
at the parameter point $\xi_{\rm ex}^{(4)}$ \eqref{eq:xi4ex}
where the attractor has two connected components
in the $\delta_L < 0$ non-invertible case.
Panel (b) uses instead $g\big(\xi_{\rm ex}^{(4)}\big)$.
}
\label{fig:att_4}
\end{figure}

Based on this we conjecture
that for any $\xi \in \cR^{(4)}_n$ with $n \ge 1$
the map \eqref{eq:BCNF2} has a chaotic attractor with exactly $2^n$ connected components,
and this is supported by the numerics in Fig.~\ref{fig:reg4}.
Fig.~\ref{fig:att_4}-a shows the attractor of \eqref{eq:BCNF2} for a typical parameter point in $\cR^{(4)}_1$, specifically
\begin{align}
    \xi_{\rm ex}^{(4)} = \left(2.2 , -0.4, -1.5, 0.4 \right).
    \label{eq:xi4ex}
\end{align}
As expected it has two connected components, one of which is contained in $\Pi_\xi$,
and both of which are affine transformations of the single-component attractor
of (1.1) with $g \big( \xi^{(4)}_{\rm ex} \big) \in \cR^{(3)}_0$ shown in Fig.~\ref{fig:att_4}-b.

\section{Reduction to one dimension}
\label{sec:reduction}

Finally we address a novelty of the non-invertible settings.
In these settings the stable eigenvalues $\lambda_L^s$ and $\lambda_R^s$ have the same sign
(they are both positive if $\delta_L > 0$ and $\delta_R < 0$,
and both negative if $\delta_L < 0$ and $\delta_R > 0$),
so it is possible for them to be equal.
From the formulas \eqref{eq:T} and \eqref{eq:D} for the points $D$ and $T$ we have
\begin{align}
    D_1 - T_1 = \frac{\lambda_L^s - \lambda_R^s}{\left(1 - \lambda_L^s\right)\left(1-\lambda_R^s\right)}.
\end{align}
Thus $D$ and $T$ coincide when the stable eigenvalues are equal.
Thus the boundaries $\phi^+(\xi) = 0$ and $\phi^-(\xi) = 0$, where $C = D$ and $C = T$ respectively, intersect where $\lambda_L^s = \lambda_R^s$,
and this is evident in Figs.~\ref{fig:reg3} and \ref{fig:reg4}.
Also for each $n \ge 1$ the boundaries $\phi^+(g^n(\xi)) = 0$ and $\phi^-(g^n(\xi)) = 0$
intersect where $\lambda_L^s = \lambda_R^s$.

\begin{figure}[h]
\begin{tabular}{cc}
  \includegraphics[scale=0.5]{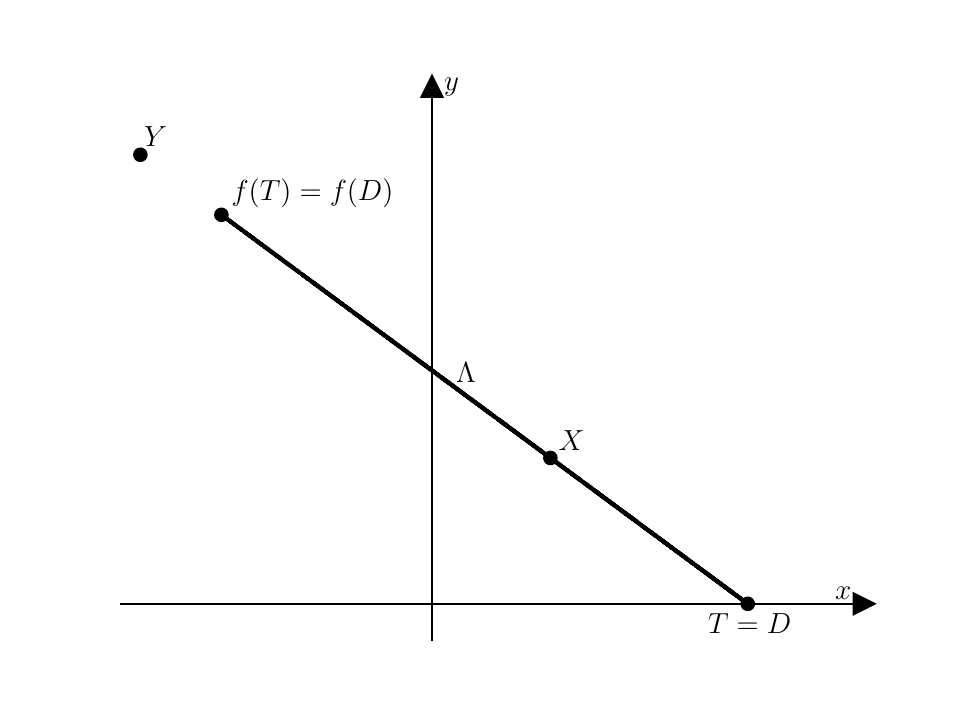} &   \includegraphics[scale=0.5]{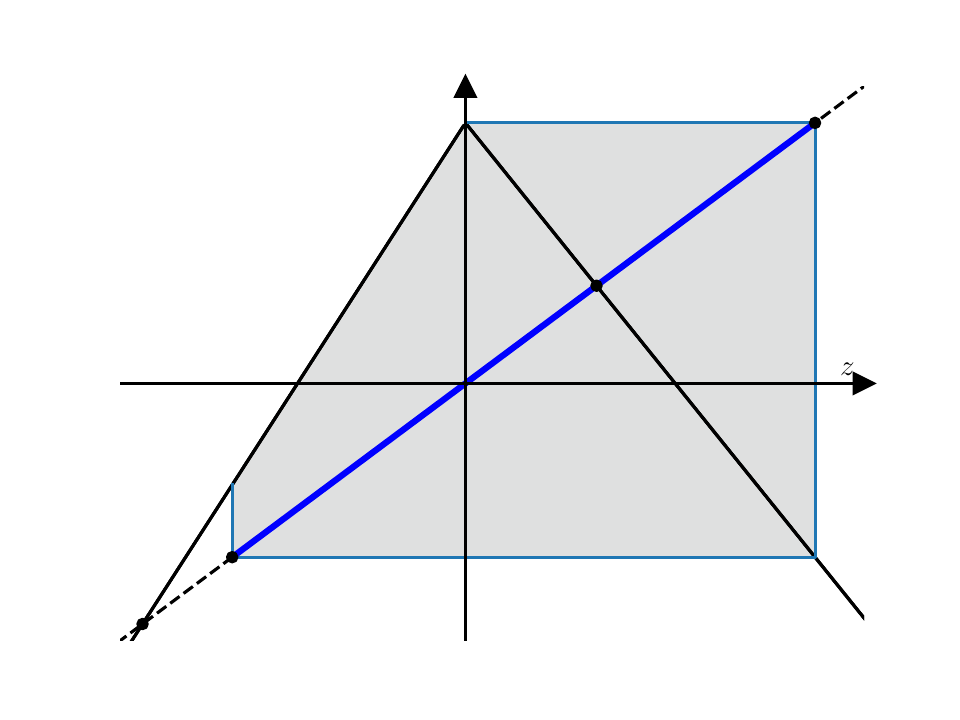}\\
(a) $\xi = \left(2.323, 0.5, -1.427, -0.4 \right)$ & (b) corresponding cobweb diagram   \\[3pt]
\end{tabular}
\caption{
Panel (a) shows the attractor of \eqref{eq:BCNF2} at the given
parameter point which belongs to $\cR^{(3)}_0$.
This attractor is the line segment from $T$ to $f_\xi(T)$.
The restriction of $f_\xi$ to the attractor is the one-dimensional
map \eqref{eq:zPrime} indicated in panel (b).
}
\label{fig:1D_att}
\end{figure}

We now show that if $\lambda_L^s = \lambda_R^s$
then the pertinent dynamics reduces to one dimension, as in
Fig.~\ref{fig:1D_att}.

\begin{proposition}
\label{pr:new_skewtent}
If $\xi \in \Phi$ with $\phi_{\rm min}(\xi) > 0$ and $\lambda_L^s = \lambda_R^s$,
then $f_\xi$ is forward invariant on the line segment from $T$ to $f_\xi(T)$.
Moreover, on this segment $f_\xi$ is conjugate to the skew tent map
\begin{equation}
z \mapsto \begin{cases}
\lambda_L^u z + 1, & z \le 0, \\
\lambda_R^u z + 1, & z \ge 0,
\end{cases}
\label{eq:zPrime}
\end{equation}
on $\left[ \lambda_R^u + 1, 1 \right]$.
\end{proposition}

\begin{proof}
The line segment from $T$ to $f_\xi(T)$ is
$$
\Gamma = \left\{ \gamma(z) \,\middle|\, z \in \big[ \lambda_R^u+1, 1 \big] \right\}, \text{~where~}
\gamma(z) = \left( \frac{z}{1-\lambda_R^s}, \frac{(1-z) \lambda_R^s}{1 - \lambda_R^s} \right),
$$
because putting $z = 1$ gives
$(x,y) = \left( \frac{1}{1-\lambda_R^s}, 0 \right) = T$ by \eqref{eq:T},
while putting $z = \lambda_R^u+1$ gives
$(x,y) = \left( \frac{1+\lambda_R^u}{1-\lambda_R^s}, \frac{-\lambda_R^s \lambda_R^u}{1-\lambda_R^s} \right)$
which is identical to $f_\xi(T) = (\tau_R T_1 + 1, -\delta_R T_1)$
(since $\tau_R = \lambda_R^s + \lambda_R^u$ and $\delta_R = \lambda_R^s \lambda_R^u$).
If $z \in [0,1]$ then
$$
f_\xi(\gamma(z)) = \left( \tau_R x + y + 1, -\delta_R x \right)
= \left( \frac{1+\lambda_R^u z}{1-\lambda_R^s}, \frac{-\lambda_R^s \lambda_R^u z}{1-\lambda_R^s} \right)
= \gamma(\lambda_R^u z + 1),
$$
while if $z \in [\lambda_R^u+1, 1]$ then
$$
f_\xi(\gamma(z)) = \left( \tau_L x + y + 1, -\delta_L x \right)
= \left( \frac{1+\lambda_L^u z}{1-\lambda_R^s}, \frac{-\lambda_R^s \lambda_L^u z}{1-\lambda_R^s} \right)
= \gamma(\lambda_L^u z + 1),
$$
and the result follows.
\end{proof}

Notice $\lambda_L^u > 1$ and $\lambda_R^u < -1$, which corresponds
in Fig.~\ref{fig:skewtent_parameterspace} to a point in some $\cR_n$, with $n \ge 0$.
Thus, with this value of $n$, the attractor of the skew tent map \eqref{eq:zPrime}
is comprised of $2^n$ disjoint intervals.
Consequently the attractor of $f_\xi$ is comprised of $2^n$ disjoint line segments.
Places where the value of $n$ changes can be computed by solving for where $\phi_0 \circ g^n = 0$ in Fig.~\ref{fig:skewtent_parameterspace}.
We computed these points numerically and have plotted them as black dots in
Figs.~\ref{fig:reg3} and \ref{fig:reg4}
from which we see that, as expected, these points are where $\phi^+ \circ g^n = 0$ and $\phi^- \circ g^n = 0$.
In this way the conjectures in \S\ref{sec:nonInvertNice} and \S\ref{sec:nonInvertNotNice} on the number of connected components of the attractors
are confirmed in the special codimension-one scenario that the stable stability multipliers
associated with $X$ and $Y$ are equal.

\section{Discussion}
\label{sec:conc}

It has long been known that symmetric tent maps and skew tent maps
readily admit attractors with $2^n$ connected components,
and that for any given slopes for the two pieces of the map the value of $n$ can be determined through renormalisation.
Here we have shown how this can be realised for two-dimensional maps
and uncovered some novel complexities relating to the possibility of stable period-two and period-four solutions. We have also shown how the attractor can be destroyed at heteroclinic bifurcations (boundary crises)
that cannot be characterised algebraically.

It remains to verify the conjectures in \S\ref{sec:or_re}--\S\ref{sec:nonInvertNotNice}
on the number of connected components of the attractor
in the various parameter regions we have defined.
We feel this should be possible by following the methodology used in \cite{GhSi22} but suspect it will be a substantial undertaking.
Chaos in higher-dimensional piecewise-linear maps
has applications to cryptography \cite{KoLi11},
and it remains to see what aspects of the renormalisation can be
described for the $N$-dimensional border-collision normal form
\cite{Si16} with no restriction on $N$.
Also it remains to see if renormalisation schemes based on other symbolic substitution rules
can be used to explain parameter regimes where \eqref{eq:BCNF2}
has attractors with other numbers of components, e.g.~three components, as described in \cite{GlSi20b}.

Finally we note that Proposition \ref{pr:new_skewtent}
provides a rare example of dimension reduction in a piecewise-smooth setting.
Dimension reduction is core element of smooth bifurcation theory 
whereby centre manifolds allow us to explain the dynamics
of high dimensional systems with low dimensional equations \cite{Ku04}.
In general the bifurcation theory of piecewise-smooth
systems is hampered by an inability to do this usually centre manifolds usually do not exist \cite{DiBu08}.
It remains to see how far Proposition \ref{pr:new_skewtent}
can be generalised to help us understand
border-collision bifurcations involving two eigenvalues that are identical, or are nearly identical.

\section*{Acknowledgements}

This work was supported by Marsden Fund contracts MAU1809 and MAU2209
managed by Royal Society Te Ap\={a}rangi.

\bibliographystyle{plain}
\bibliography{main.bib}

\end{document}